\documentclass[twocolumn, twocolappendix, appendixfloatss]{aastex631}
\makeatletter
\let\frontmatter@title@above=\relax
\makeatother

\let\tablenum\relax
\usepackage{siunitx}
\usepackage{lipsum}

\date{August 31, 2022}

\shorttitle{CH+ electron recombination}
\shortauthors{Paul et al.}

\begin{document}
	
\title{Experimental determination of the dissociative recombination rate coefficient for rotationally-cold CH$^{+}$ and its implications for the diffuse cloud chemistry}

\correspondingauthor{Daniel Paul}
\email{daniel.r.paul@t-online.de}

\author[0000-0001-7625-4398]{Daniel Paul}
\affiliation{Max-Planck-Institut für Kernphysik, 69117 Heidelberg, Germany}
\affiliation{Columbia Astrophysics Laboratory, Columbia University, New York, NY 10027, USA}

\author[0000-0002-5066-2550]{Manfred Grieser}
\affiliation{Max-Planck-Institut für Kernphysik, 69117 Heidelberg, Germany}

\author{Florian Grussie}
\affiliation{Max-Planck-Institut für Kernphysik, 69117 Heidelberg, Germany}

\author{Robert von Hahn}
\affiliation{Max-Planck-Institut für Kernphysik, 69117 Heidelberg, Germany}

\author[0000-0002-3571-5765]{Leonard W. Isberner}
\affiliation{I. Physikalisches Institut, Justus-Liebig-Universität Gießen, 35392 Gießen, Germany}
\affiliation{Max-Planck-Institut für Kernphysik, 69117 Heidelberg, Germany}

\author[0000-0003-3782-0814]{\'{A}bel K\'{a}losi}
\affiliation{Max-Planck-Institut für Kernphysik, 69117 Heidelberg, Germany}
\affiliation{Columbia Astrophysics Laboratory, Columbia University, New York, NY 10027, USA}

\author[0000-0003-1727-8319]{Claude Krantz}
\affiliation{Max-Planck-Institut für Kernphysik, 69117 Heidelberg, Germany}

\author[0000-0003-0511-0738]{Holger Kreckel}
\affiliation{Max-Planck-Institut für Kernphysik, 69117 Heidelberg, Germany}

\author[0000-0003-3391-6886]{Damian Müll}
\affiliation{Max-Planck-Institut für Kernphysik, 69117 Heidelberg, Germany}

\author[0000-0001-8341-1646]{David A. Neufeld}
\affiliation{Department of Physics \& Astronomy, Johns Hopkins University, Baltimore, MD 21218, USA}

\author[0000-0002-1111-6610]{Daniel W. Savin}
\affiliation{Columbia Astrophysics Laboratory, Columbia University, New York, NY 10027, USA}

\author[0000-0002-6166-7138]{Stefan Schippers}
\affiliation{I. Physikalisches Institut, Justus-Liebig-Universität Gießen, 35392 Gießen, Germany}

\author[0000-0002-5020-3880]{Patrick Wilhelm}
\affiliation{Max-Planck-Institut für Kernphysik, 69117 Heidelberg, Germany}

\author[0000-0003-1198-9013]{Andreas Wolf}
\affiliation{Max-Planck-Institut für Kernphysik, 69117 Heidelberg, Germany}

\author[0000-0003-0030-9510]{Mark G. Wolfire}
\affiliation{Department of Astronomy, University of Maryland, College Park, MD 20742, USA}

\author[0000-0003-2520-343X]{Old\v{r}ich Novotn\'{y}}
\affiliation{Max-Planck-Institut für Kernphysik, 69117 Heidelberg, Germany}




\begin{abstract}
	
Observations of CH$^+$ are used to trace the physical properties of diffuse clouds, but this requires an accurate understanding of the underlying CH$^+$ chemistry. Until this work, the most uncertain reaction in that chemistry was dissociative recombination (DR) of CH$^+$. Using an electron-ion merged-beams experiment at the Cryogenic Storage Ring, we have determined the DR rate coefficient of the CH$^+$ electronic, vibrational, and rotational ground state applicable for different diffuse cloud conditions. Our results reduce the previously unrecognized order-of-magnitude uncertainty in the CH$^+$ DR rate coefficient to $\sim \pm 20\%$ and are applicable at all temperatures relevant to diffuse clouds, ranging from quiescent gas to gas locally heated by processes such as shocks and turbulence. Based on a simple chemical network, we find that DR can be an important destruction mechanism at temperatures relevant to quiescent gas. As the temperature increases locally, DR can continue to be important up to temperatures of $ \sim\SI{600}{K} $ if there is also a corresponding increase in the electron fraction of the gas. Our new CH$^+$ DR rate coefficient data will increase the reliability of future studies of diffuse cloud physical properties via CH$^+$ abundance observations.
	
\end{abstract}

\keywords{\href{http://astrothesaurus.org/uat/381}{Diffuse molecular clouds (381)},
	\href{http://astrothesaurus.org/uat/2004}{Laboratory astrophysics (2004)}, \href{http://astrothesaurus.org/uat/2081}{Reaction rates (2081)},
	\href{http://astrothesaurus.org/uat/2262}{Electron recombination reactions (2262)}
}


\makeatletter

\renewcommand\@makefntext[1]{%
	\setlength\parindent{0em}%
	\setlength\rightskip{0.2em}
	\mbox{\textsuperscript{\@thefnmark}~}{#1}}

\makeatother

\section{Introduction} \label{sec:intro}

The methylidine cation $ \mathrm{CH}^+ $ was the first molecular ion detected in the interstellar medium \citep{Douglas1941}. In the cold and dilute environment of diffuse clouds it radiatively relaxes to an internal temperature of $ T_\mathrm{ex}\sim\SI{3}{K} $ and is predominantly found in its vibrational $ v=0 $ and rotational $J=0$ ground state \citep{Godard2013}.
	
Despite its early detection, the high abundance of $ \mathrm{CH}^+ $ in diffuse clouds still remains an astrochemical puzzle \citep{Indriolo2010}. This is because the only known production mechanism via C$ ^+ $,
\begin{equation}
	\mathrm{C}^+ + \mathrm{H_2} \rightarrow \mathrm{CH}^+ + \mathrm{H} , \label{eq:production}
\end{equation}
is endothermic by $ \sim\SI{0.4}{eV} $ \citep{Gerlich1987} and cannot proceed at quiescent diffuse cloud  kinetic temperatures\footnote[1]{We distinguish between the kinetic temperature $T_\mathrm{k}$ of the gas in the cloud, which characterizes the velocity distributions for all particles, and the excitation temperature $T_\mathrm{ex}$, which describes the internal state-population distribution of the CH$^+$ molecules. The kinetic temperature is sometimes also called the translational temperature.} of $ T_\mathrm{k}=\SI[parse-numbers=false]{40-130}{K}$ ($k_\mathrm{B}T_\mathrm{k}=\SI[parse-numbers=false]{3-11}{meV} $, where $k_\mathrm{B}$ is the Boltzmann constant; \citealt{Shull2021})\footnote[2]{In principle, CH$^+$ can also form via
\begin{displaymath}
	\mathrm{C}^{2+} + \mathrm{H_2} \rightarrow \mathrm{CH}^+ + \mathrm{H}^+, \label{eq:production2+}
\end{displaymath}
but this reaction is negligible because of its experimentally-known low rate coefficient of $ \lesssim\SI{2e-12}{cm^3\,s^{-1}} $ and the low abundance of C$ ^{2+}$ compared to C$^+$ in diffuse clouds (\citealt{Plasil2021} and references therein).}. 

A number of mechanisms have been proposed to address the CH$^+$ abundance puzzle, which drive Equation (\ref{eq:production}) by locally heating the gas or by producing a relative velocity between ions and neutrals that increase the effective temperature of the reaction. Such mechanisms include: Alfv\'{e}n waves \citep{Federman1996}, turbulent mixing \citep{Lesaffre2007, Valdivia2017}, magnetohydrodynamic shocks \citep{Lesaffre2013}, turbulent dissipation in vortices \citep{Godard2014}, ion-neutral drift \citep{Valdivia2017, Moseley2021}, and magnetohydrodynamic turbulence in clouds \citep{Moseley2021}. These processes are hypothesized to locally heat the gas or raise the effective temperature to greater than $ \SI{200}{K} $ ($k_\mathrm{B}T_\mathrm{k}=\SI{17}{meV}$) and predict CH$^+$ abundances that can differ from one another by several orders of magnitude. One aim of our work here is to reduce the uncertainties in the CH$^+$ chemical kinetics data so that remaining discrepancies between models and observations can be more reliably used to constrain the astrophysical properties of the diffuse gas and not be attributed to a lack of understanding of the underlying chemical kinetics.

Using $ \mathrm{CH}^+ $ observations to constrain which of these mechanisms is driving the observed abundances, requires an accurate understanding of the underlying $ \mathrm{CH}^+ $ chemistry. Reaction (\ref{eq:production}) has been studied both experimentally \citep{Hierl1997} and theoretically \citep{Wu2021}. The dominant destruction reactions are \citep{Myers2015}
\begin{eqnarray}
	\mathrm{CH}^+ + \mathrm{H}\ \ \, &\rightarrow&\ \mathrm{C}^+ + \mathrm{H_2} , \label{eq:destructionH}\\
	\mathrm{CH}^+ + \mathrm{H_2}\ &\rightarrow&\ \mathrm{CH_2}^+ + \mathrm{H} , \label{eq:destructionH2}\\
	\mathrm{CH}^+ + \mathrm{e^-}\ &\rightarrow&\ \mathrm{C} + \mathrm{H} . \label{eq:dr}
\end{eqnarray}
The rate coefficients for hydrogen abstraction via Reaction (\ref{eq:destructionH}) and dissociative recombination (DR) in Reaction (\ref{eq:dr}) have been calculated theoretically (e.g., \citealt{Faure2017}). We are unaware of any theoretical calculations for Reaction (\ref{eq:destructionH2}) more sophisticated than the Langevin value \citep{Kim1975}. Experimentally, \citet{Plasil2011} and \citet{Gerlich2011} measured the rate coefficients for Reaction (\ref{eq:destructionH}) at $ \SI[parse-numbers=false]{12-100}{K} $ and Reaction (\ref{eq:destructionH2}) at $ \SI{50}{K} $, respectively, approximating diffuse cloud environments. However, we are unaware of any cryogenic measurement for the DR process, Reaction (\ref{eq:dr}).

In diffuse clouds, DR represents the largest uncertainty among all chemical reactions that form and destroy $ \mathrm{CH}^+ $. Calculations of the DR cross section are difficult due to the large number of intermediate states involved in the dissociation dynamics \citep{Chakrabarti2018,Mezei2019}. Experimentally, single-pass merged-beams experiments have been performed by \citet{Mitchell1978}, where the ions were likely to be electronically, vibrationally, and rotationally excited. The only DR rate-coefficient results for electronically and vibrationally relaxed ($ v=0 $) ions were obtained at the room-temperature Test Storage Ring (TSR; \citealt{Amitay1996}). However, the energy resolution in this measurement was $ \sim \SI{20}{meV} $, i.e., much higher than the collision energies in quiescent gas of a diffuse cloud. Additionally, the ions occupied a broad range of rotational states with only $\sim6\%$ in the $J=0$ state. Given that recent experimental work has found that the $J$-specific DR rate coefficient can vary by over an order of magnitude \citep{Novotny2019}, the uncertainty in the experimentally derived CH$^+$ rate coefficient from the above two CH$^+$ works is likely to be a previously unrecognized order of magnitude. This makes the DR process the most uncertain reaction in the chemistry forming and destroying $ \mathrm{CH}^+ $ in diffuse clouds.

With the development of the Cryogenic Storage Ring (CSR; \citealt{vonHahn2016}), it has become possible to perform electron-ion merged-beams collision studies with internally cold molecular ions. Recent CSR DR experiments with $ \mathrm{HeH}^+ $ revealed a strong $ J $-dependence of the rate coefficient \citep{Novotny2019}, demonstrating the importance of probing rotationally cold ions. In this work, we use the CSR to create an electronically and rovibrationally cold ($ v=0 $, $ J=0-2 $) ensemble of $ \mathrm{CH}^+ $ molecules, measure DR with an order-of-magnitude improved collision energy resolution of $ \sim\SI{2}{meV} $, and extract a rate coefficient for the rovibrational $ \mathrm{CH}^+ $ ground state appropriate for diffuse cloud conditions, thus refining the most uncertain reaction in the $ \mathrm{CH}^+ $ diffuse cloud chemistry.

\section{Experimental} \label{sec:exp}

\begin{figure*}[]
	\epsscale{1}
	\plotone{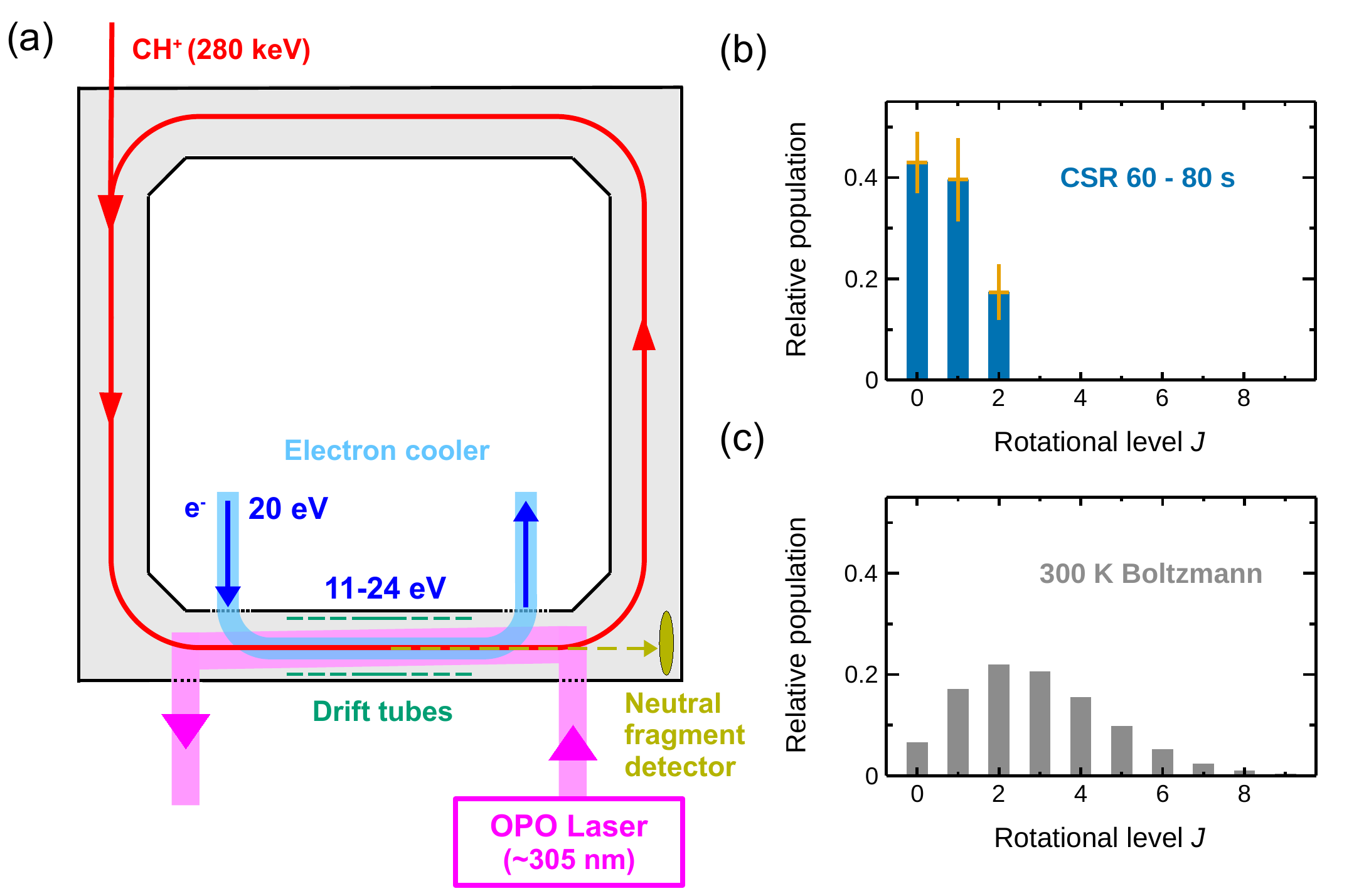}
	\caption{(a) Schematic of the CSR experimental setup for DR measurements of CH$^+$  and probing of its internal level populations. See text for details. (b) Measured relative $ J $-level populations in the CSR for storage times of $ \SI[parse-numbers=false]{60-80}{s} $ with statistical one-sigma error bars. (c) Relative $ J $-level populations predicted in TSR for a thermal equilibrium at $ T=\SI{300}{K} $.}
	\label{fig:setup}
\end{figure*}

In order to determine the DR rate coefficient of the $ X^1\Sigma^+(v=0,J=0) $ ground state of $ \mathrm{CH}^+ $, we have used the experimental setup sketched in Figure \ref{fig:setup}(a). We stored and phase-space cooled a $ \mathrm{CH}^+ $ ion beam inside the CSR, prepared internally cold ions, utilized \emph{in situ} diagnostic methods to follow time-dependent changes of the internal level populations, and finally measured DR with well-defined mixtures of only a few $ J $ levels to infer the rate coefficient for the $ \mathrm{CH}^+ $ ground electronic, vibrational, and rotational state. 

Phase-space cooling\footnote[3]{Commonly also called electron cooling in literature} of the $ \mathrm{CH}^+ $ ion beam was applied to reduce its diameter and energy spread \citep{Poth1990}. The stored ion beam was overlapped with a larger diameter, nearly mono-energetic electron beam in the CSR electron cooler section (Figure \ref{fig:setup}(a)). Phase-space cooling was reached by matching the electron and ion beam positions and velocities for the first $ \SI{21}{s} $ of storage, before the DR measurements began. Phase-space cooling was also regularly applied during the collision experiments to maintain the beneficial ion-beam properties.

Internal cooling of the initial electronically and rovibrationally hot ion beam, was achieved by radiative decay and by inelastic collisions with the merged, nearly monoenergetic, velocity-matched electron beam. Radiative cooling proceeded by interactions with the $ \sim\SI{20}{K} $ two-component blackbody radiation field in the CSR \citep{Kalosi2022}. This dominated the initial internal cooling of highly excited states. After $ \SI{21}{s} $ of storage, most ions were in their ground electronic state $ X^1\Sigma^+ $ with a rovibrationally cold distribution of $ v=0$ and $J=0-2 $. Further rotational cooling was driven by inelastic electron collisions and advanced towards an effective internal ion temperature of $ \sim\SI{26}{K} $ \citep{Kalosi2022}.

\begin{figure*}[]
	\epsscale{0.8}
	\plotone{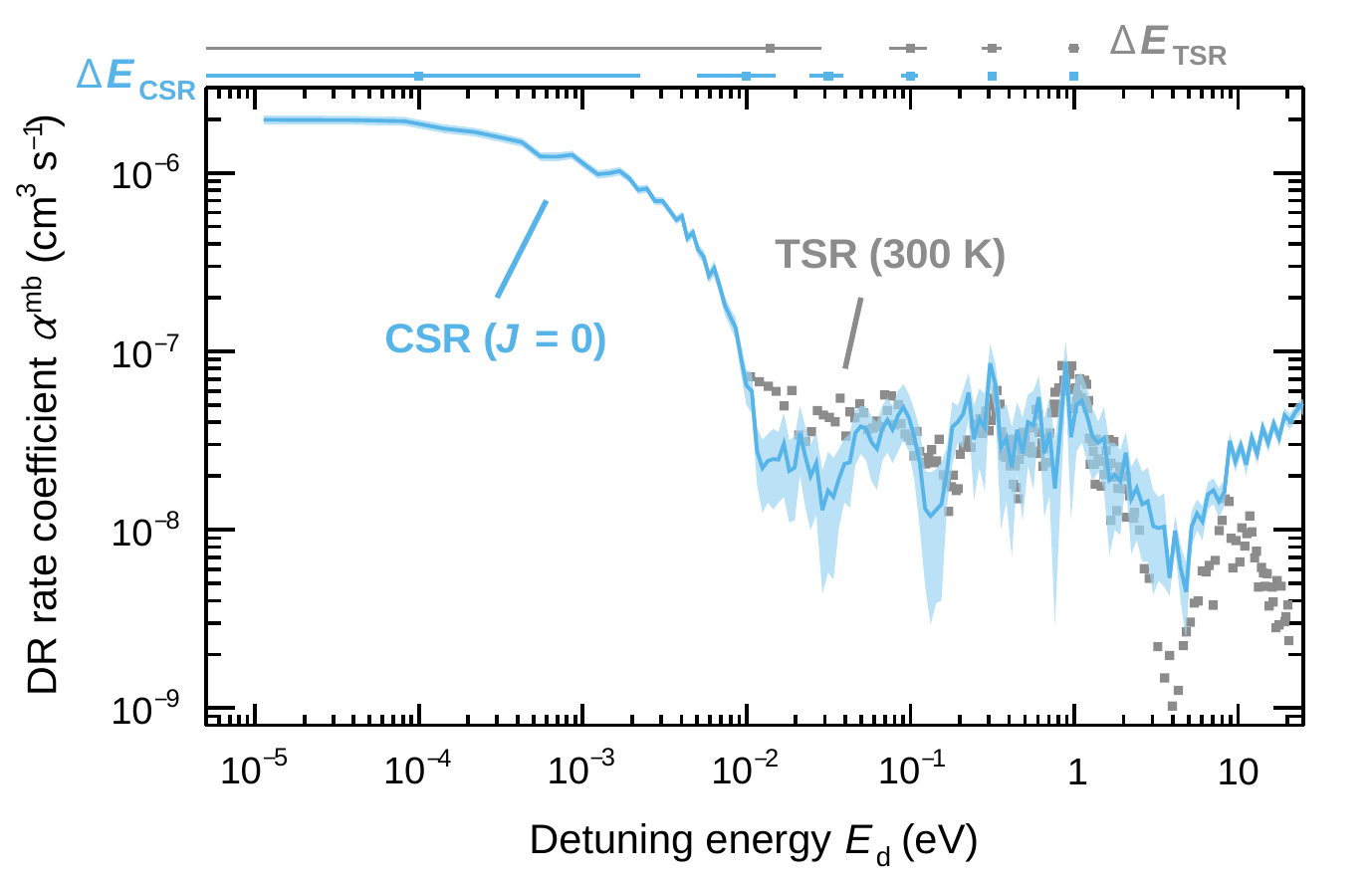}
	\caption{Merged-beams DR rate coefficient (blue line) for the CH$^{+}$ ground state $X^1\Sigma^+ (v=0,J=0)$ with the statistical one-sigma uncertainties (blue shaded area). The one-sigma uncertainty of the absolute scale is $ \pm13 \% $. The room-temperature TSR results of \citet{Amitay1996}, which have a $ \pm50 \% $ absolute scaling systematic uncertainty, are shown by the gray squares. The symbols and lines on top of the figure indicate the full width at half maximum (FWHM) energy resolution $ \Delta E $ of the CSR and TSR measurements. The divergence of the CSR from the TSR data for $ E_\mathrm{d}\gtrsim\SI{10}{eV} $ is due to contribution from dissociative excitation (see text). The data behind the TSR and CSR graphs are both available. The latter are part of Figure \ref{fig:ratecoeffmeas}(b).}
	\label{fig:ratecoeff}
\end{figure*}

We monitored the internal cooling \emph{in situ} by probing $ J $-level-resolved photodissociation of $ \mathrm{CH}^+ $ with an optical parametric oscillator (OPO) laser at a tunable wavelength near $ \sim\SI{305}{nm} $ and measuring the resulting H atoms on a neutral-fragment detector (Figure \ref{fig:setup}(a); \citealt{Kalosi2022}). A combined radiative and collisional equilibrium was reached by $ t\sim\SI{70}{s} $, with the $ J=0 $ and $ J=1 $ levels dominating the rotational populations (Figure \ref{fig:setup}(b)). In contrast to the room-temperature TSR rotational distribution (Figure \ref{fig:setup}(c)), in the present work over $ 40\% $ of the ions populate the $J=0 $ level. In addition to the $ X^1\Sigma^+ $ ground state, there is a small fraction of ions ($ <13\% $ at $ \SI{21}{s} $) in the metastable $ a^3\Pi $ state, which decays with a lifetime of $ \sim\SI{10}{s} $ and which we monitored by its unique DR fragmentation pattern. In this way, we obtained full knowledge about the time-dependent populations of all contributing rotational levels ($ p_J(t) [J=0-2] $) and the metastable state ($ p_\mathrm{m}(t) $), see Appendix \ref{app:rotational}.

DR rate-coefficient measurements on the internally cold $ \mathrm{CH}^+ $ ions used the electron beam from the electron cooler as a merged-beams collision target. Neutral C and H atoms resulting from DR in the interaction region of the electron cooler were collected downstream by our neutral-fragment detector. From the measured count rate we determined the merged-beams DR rate coefficient $ \alpha^\mathrm{mb} $, related to the DR cross section $\sigma$ via
\begin{equation} \label{eq:convolutionShort}
	\alpha^\mathrm{mb}=\langle \sigma v_\mathrm{c}  \rangle .
\end{equation}
Here, $\sigma(E)$ and the collision velocity $ v_\mathrm{c}(E) $ are functions of the collision energy $ E $ and $\langle...\rangle $ designates the mean value over the full distribution of $ v_\mathrm{c} $  inside the electron-ion overlap region. This distribution depends on the transverse and longitudinal temperatures in the rest frame of the electron beam. We achieved $ k_\mathrm{B}T_\perp=\SI[parse-numbers=false]{2.0^{+1.0}_{-0.5}}{meV} $ and $ k_\mathrm{B}T_\parallel\sim\SI{0.5}{meV} $, respectively, corresponding to a collision energy resolution of $ \sim\SI{2}{meV} $ at matched velocities. Here and throughout, all uncertainties are quoted at an one-sigma confidence level with $ ^{+}_{-}$ for asymmetric uncertainties and numbers in parenthesis for symmetric uncertainties.

In order to study the energy dependence of the DR process, electron energies were scanned by adjusting the potential of several drift tubes in the merged-beams interaction section (Figure \ref{fig:setup}(a)). This yielded $ \alpha^\mathrm{mb}(E_\mathrm{d}) $, where the detuning energy $ E_\mathrm{d} $ is defined as the collision energy of ideal, mono-energetic electron and ion beams, given by the drift-tube potentials. Details on the electron-ion collision setup and on the absolute scaling of $ \alpha^\mathrm{mb}(E_\mathrm{d}) $ are given in Appendices \ref{app:collSetup} and \ref{app:absRate}, respectively.

The astrophysically relevant $ \mathrm{CH}^+ $ DR rate coefficient was inferred from our storage-time-dependent measurements of $ \alpha^\mathrm{mb}(E_\mathrm{d},t) $. For storage times of $ t=\SI[parse-numbers=false]{21-100}{s} $, we used the experimentally determined internal level populations to extract the merged-beams rate coefficient $ \alpha^{\mathrm{mb}}_{J=0}(E_\mathrm{d}) $ for the pure $ J=0 $ state (Appendix \ref{app:stateResolved}). From this, we generated the DR kinetic rate coefficient $ \alpha^\mathrm{k}_{J=0}(T_\mathrm{k}) $ for the $ J=0 $ state (Appendix \ref{app:deconvolution}), needed for diffuse cloud chemistry at a gas kinetic temperature $ T_\mathrm{k} $.

\newpage

\section{Results} \label{sec:res}

\subsection{Merged-beams rate coefficient}

The rate coefficient $\alpha^{\mathrm{mb}}_{J=0}(E_\mathrm{d}) $ in Figure \ref{fig:ratecoeff} decays by two orders of magnitude in the energy range from $ \SI[parse-numbers=false]{10^{-5}}{eV} $ to $ \SI[parse-numbers=false]{10^{-2}}{eV} $ and shows various resonances in the region of $ \SI[parse-numbers=false]{0.03-2}{eV} $. These resonances can be attributed to capture into neutral Rydberg states of the CH molecule and subsequent coupling to a dissociative state (\citealt{Amitay1996}, \citealt{Mezei2019}). At energies beyond $ \sim\SI[parse-numbers=false]{3.6}{eV} $, a non-resonant DR channel opens up and leads to an increased DR rate coefficient (\citealt{Amitay1996}; \citealt{Chakrabarti2018}). In addition, $ \mathrm{CH}^+ $ dissociative excitation (DE) sets in at a threshold of $ \sim\SI{3.6}{eV} $ (\citealt{Bannister2003}; \citealt{Chakrabarti2017}), which we cannot distinguish from DR owing to our experimental setup. 

Comparing our data to the room-temperature TSR data of \citet{Amitay1996}, which had rotational populations close to the simulation shown in Figure \ref{fig:setup}(c), we obtain information on the DR rotational level dependence at $ E_\mathrm{d}>\SI[parse-numbers=false]{0.03}{eV} $ and on measurement-related effects. At $ \SI[parse-numbers=false]{0.03-2}{eV} $, the resonant structures and absolute values agree well between the two data sets, despite the differing $ J $-level populations. This indicates the absence of major $ J $-dependent resonances at detuning energies above $ \SI[parse-numbers=false]{0.03}{eV} $. This conclusion is corroborated by the close match of the CSR results for the $ J=0 $ and $ J=1 $ merged-beams rate coefficients for $ E_\mathrm{d}>\SI{0.03}{eV} $ (Appendix \ref{app:stateResolved}). At $ E_\mathrm{d}\gtrsim\SI[parse-numbers=false]{3.6}{eV} $, due to a measurement-related effect, our data are enhanced compared to the TSR results, where it was possible to discriminate between DR and DE. In addition, our collision geometry imposes a rate coefficient baseline of $ \sim\SI{6e-9}{cm^3\,s^{-1}} $ to the data, which becomes visible at $ E_\mathrm{d}>\SI[parse-numbers=false]{2}{eV} $ (Appendix \ref{app:absRate}). Both effects are accounted for in the determination of the $ \mathrm{CH}^+ $ DR kinetic temperature rate coefficient (Appendix \ref{app:deconvolution}).

Due to the improved energy resolution of the CSR depicted on top of Figure \ref{fig:ratecoeff}, we were able to probe the diffuse cloud collision-energy regime ($\SI[parse-numbers=false]{0.001-0.02}{eV}$), where no experimental data have previously been available. Our results reveal a remarkable increase in the DR rate coefficient towards lower detuning energies. This feature is considerably stronger than the $ E_\mathrm{d}^{-1/2} $ behavior that would result from a typical $ E^{-1} $ cross section dependence of non-resonant DR \citep{Mitchell1990}. This can be attributed either to a constructive resonance near $ E_\mathrm{d}=\SI{0}{eV} $ or a destructive resonance near $ E_\mathrm{d}=\SI{0.01}{eV} $. We also note that at $ E_\mathrm{d}=\SI{0.012}{eV} $ the $ J=0 $ rate coefficient is lower by a factor of $ 5.4^{+5.0}_{-2.0}$ compared to $ J=1 $. However, we find no significant deviation for $ E_\mathrm{d}<\SI{0.008}{eV} $ and $ E_\mathrm{d}>\SI{0.02}{eV} $ (Appendix \ref{app:stateResolved}). 

\begin{figure*}[]
	\epsscale{1.15}
	\plotone{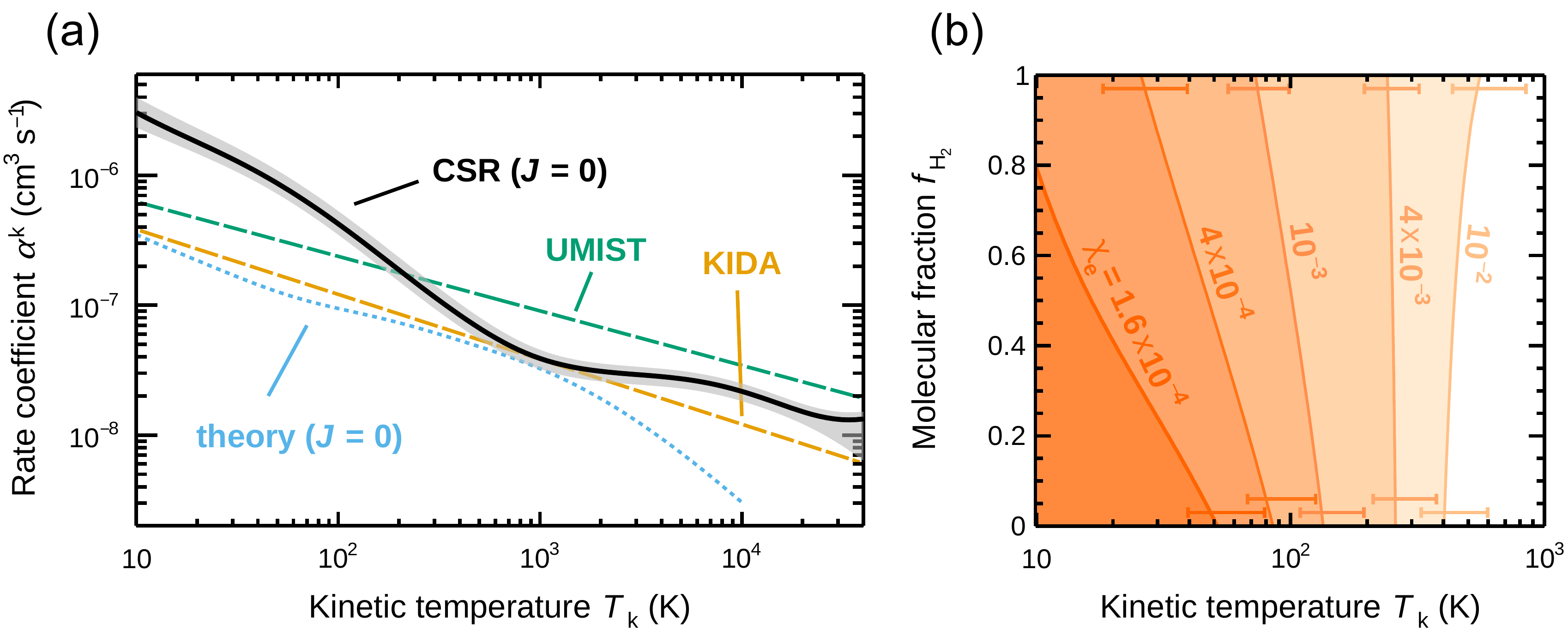}
	\caption{CH$^{+}$ DR kinetic temperature rate coefficient and its relevance for the CH$^{+}$ destruction in diffuse interstellar clouds. (a) The thick solid line shows the experimentally determined $ J=0 $ kinetic temperature rate coefficient of this work. The one-sigma error band in gray represents the quadrature sum of all systematic uncertainties.  The dashed lines indicate rate coefficients from the UMIST (\citealt{McElroy2013}, upper line) and KIDA (\citealt{Wakelam2012}, lower line) astrochemistry databases. The calculated $ J=0 $ rate coefficient of \citet{Mezei2019} is drawn as dotted line. (b) Diffuse cloud parameter ranges where DR is the dominant CH$^+$ destruction process (colored) for a standard electron fraction ($ \chi_{\mathrm{e}}=\SI{1.6e-4}{} $) and for enhanced values of $ \chi_{\mathrm{e}} $ as given. One-sigma error bars for $ f_\mathrm{H_2}=1 $ and $ f_\mathrm{H_2}=0 $ are shown at the top and bottom of the figure, respectively. They were calculated from the combined systematic uncertainties for the rate coefficients $ \alpha_\mathrm{H} $ ($\sim50\%$, \cite{Plasil2011}), $ \alpha_\mathrm{H_2} $ ($\sim20\%$, \cite{Gerlich2011}), and $ \alpha^\mathrm{k}_{J=0} $ ($ \sim20\% $, this work).}
	\label{fig:plasmarate}
\end{figure*}

\subsection{Kinetic temperature rate coefficient}

To obtain the DR kinetic temperature rate coefficient $ \alpha^\mathrm{k}_{J=0}(T_\mathrm{k}) $ for the pure $ J=0 $ state, we converted the $\alpha^{\mathrm{mb}}_{J=0}(E_\mathrm{d}) $ data (Figure \ref{fig:ratecoeff}) as described in Appendix \ref{app:deconvolution}. The result in Figure \ref{fig:plasmarate}(a) shows that our improved experimental energy resolution enables us to reliably determine the kinetic rate coefficient down to $ T_\mathrm{k}\sim\SI{10}{K} $. With the broad range of collision energies probed, we can provide data up to $ T_\mathrm{k}\sim\SI{40000}{K} $, close to the temperature equivalent of the dissociation threshold $ D_\mathrm{0}/k_\mathrm{B}\sim\SI{47000}{K} $ \citep{Hechtfischer2002}. The kinetic rate-coefficient shape can be approximated by a $ T_\mathrm{k}^{-0.5} $ behavior with two broad peaks at low and high temperatures of $ T_\mathrm{k}\sim10 $ and $ \sim\SI{8000}{K} $, respectively.  These can be attributed to the maxima in the merged-beams rate coefficient at $ E_\mathrm{d}\lesssim\SI{0.005}{eV} $ and $ E_\mathrm{d}=\SI[parse-numbers=false]{0.2-2}{eV} $ in Figure \ref{fig:ratecoeff}, respectively. A discussion about the uncertainties displayed as an error band is presented in Appendix \ref{app:deconvolution}.

Figure \ref{fig:plasmarate}(a) compares our experimental results to values in the UMIST \citep{McElroy2013} and KIDA \citep{Wakelam2012} astrochemistry databases. Here, the UMIST entry cites a review by \citet{Mitchell1990} reporting the results of a single-pass merged-beams experiment by \citet{Mitchell1978}, where multiple electronic, vibrational, and rotational states were likely to be involved. Thus, those results are not expected to represent diffuse cloud conditions. The KIDA rate coefficient is referred to as estimated. Both database rate coefficients follow a simple behavior near $ T_\mathrm{k}^{-0.5} $ and clearly miss the temperature-dependent features seen in our data. In the temperature regime for quiescent gas in diffuse clouds, the new CSR results exceed the database values by up to a factor of 6. Conversely, for gas heated by shocks or turbulence, our data are up to a factor of $2.5$ lower.

We also provide a comparison to multichannel quantum defect theory calculations for the CH$^{+} (J=0) $ DR cross section \citep{Mezei2019}, which we converted into a kinetic rate coefficient as shown in Appendix \ref{app:deconvolution}. Compared to their results, our data are significantly larger at low temperatures. In particular, in the temperature regime of diffuse clouds ($ T_\mathrm{k}=\SI[parse-numbers=false]{40-130}{K} $) theory underestimates the DR rate coefficient by up to a factor of $ 7 $. For heated gas at $ T_\mathrm{k}\sim\SI{1000}{K} $, the two results match but begin to diverge again for $ T_\mathrm{k}\gtrsim\SI{3000}{K} $, where our data exceed theory by up to a factor of 7.

\section{Astrochemical implications} \label{sec:astro}

Understanding the CH$^{+}$ abundance in diffuse clouds has been limited, in part, by uncertainties in the relevant chemistry. Our measurements of the CH$^{+}(J=0)$ DR rate coefficient reduce its uncertainty from above an order of magnitude to $ \sim20\% $ and thus provide a significant improvement in the reliability of the CH$^{+}$ chemistry. Remaining discrepancies between observations and models can now be more solidly used to constrain the physical properties of these clouds. In Appendix \ref{app:plasmaRateResults} we provide analytical representations of our kinetic temperature rate coefficient results for ready incorporation into astrochemical databases.

Based on our measured DR rate coefficient, we can determine the relative importance of the CH$^{+}$ destruction mechanisms given by Reactions (\ref{eq:destructionH})-(\ref{eq:dr}). A temperature-dependent rate coefficient $ \alpha_\mathrm{H}(T_\mathrm{k}) $ for Reaction (\ref{eq:destructionH}) with pure CH$^{+} (J=0)$ ions was derived from experimental data by \citet{Plasil2011} (Equation (6a) of their work), which we extrapolate up to $ T_\mathrm{k}=\SI{600}{K} $. For CH$^{+}$ destruction by molecular hydrogen (Reaction (\ref{eq:destructionH2})), we use the rate coefficient $ \alpha_\mathrm{H_2}=\SI{1.2e-9}{cm^3\,s^{-1}} $ from  \citet{Smith1977} and \citet{Gerlich2011}, which were obtained for fully thermal temperatures, i.e., gas kinetic and excitation temperatures, of $T_\mathrm{k}=T_\mathrm{ex}=\SI{300}{} $ and $ \SI{50}{K} $, respectively. Given the match of the rate coefficients at different $ T_\mathrm{k} $, we assume that $ \alpha_\mathrm{H_2}(\SI{10}{K}\leq T_\mathrm{k}\leq\SI{600}{K})=\SI{1.2e-9}{cm^3\,s^{-1}}$.

To compare destruction by atomic and molecular hydrogen collisions to destruction by DR, we define a critical DR rate coefficient $ \alpha_\mathrm{cr,DR} $ at which the CH$^{+}$ destruction by both processes is balanced. Diffuse clouds are characterized by their electron densities $ n_\mathrm{e} $, their atomic and molecular hydrogen number densities, $n(\mathrm{H})$ and $n(\mathrm{H_2})$, and related quantities: the hydrogen nucleus number density $ n_\mathrm{H}=n(\mathrm{H})+2n(\mathrm{H_2}) $, the molecular fraction $f_\mathrm{H_2}=2n(\mathrm{H_2})/n_\mathrm{H}$, and the electron fraction $ \chi_\mathrm{e}=n_\mathrm{e}/n_\mathrm{H}$. The critical DR rate coefficient is then given by
\begin{equation} \label{eq:alphaCr}
	\alpha_\mathrm{cr,DR}(T_\mathrm{k})=\frac{(1-f_\mathrm{H_2})\alpha_\mathrm{H}(T_\mathrm{k})+f_\mathrm{H_2}\alpha_\mathrm{H_2}(T_\mathrm{k})/2}{\chi_\mathrm{e}}.
\end{equation}
Destruction by DR dominates over that by hydrogen collisions if $ \alpha^\mathrm{k}_{J=0}>\alpha_\mathrm{cr,DR} $.

Combining Equation (\ref{eq:alphaCr}) with the condition $ \alpha^\mathrm{k}_{J=0}>\alpha_\mathrm{cr,DR} $, we calculated the diffuse cloud parameters ($ T_\mathrm{k} $, $ f_\mathrm{H_2}$), at which CH$^{+}$ destruction is mainly due to DR. The result is presented in Figure \ref{fig:plasmarate}(b). Assuming that $ \chi_\mathrm{e}=\SI{1.6e-4}{} $ matches the relative C$ ^+ $ abundance \citep{Sofia2004}, we find that for cold and mainly atomic gas ($ T_\mathrm{k}\lesssim\SI{50}{K} $, $ f_\mathrm{H_2}\lesssim0.05 $) DR is the dominant destruction process. Above this temperature, hydrogen abstraction becomes dominant. In case of higher molecular fractions ($ f_\mathrm{H_2}\gtrsim0.1 $), we find that for gas temperatures of $ T_\mathrm{k}>\SI{40}{K} $ destruction by DR is negligible compared to hydrogen abstraction. Given the assumptions made for $ \alpha_\mathrm{H} $ and $ \alpha_\mathrm{H_2} $, the uncertainties displayed in Figure \ref{fig:plasmarate}(b) are likely underestimated. For example, no lower limit for the $ J=0 $ rate coefficient $ \alpha_\mathrm{H} $ could be determined by \citet{Plasil2011}, which might lead to a higher relevance of DR in fully atomic clouds, beyond the results of Figure \ref{fig:plasmarate}(b).

The results in Figure \ref{fig:plasmarate}(b) can be put into the context of the CH$^{+}$ abundance puzzle in diffuse clouds. In quiescent gas the increased DR kinetic rate coefficient determined in our work leads to a factor of $\sim2$ enhancement in the destruction of CH$^{+}$ than previously assumed. However, because of the local heating mechanisms required to produce CH$^{+}$ and the high reactivity of the molecules, which prevents them from surviving long enough to flow into cold regions of the cloud, very little CH$^{+}$ abundance is expected in cold gas. Instead, the CH$^{+}$ chemistry is expected to take place at $ T_\mathrm{k}\gtrsim\SI{1000}{K} $ \citep{Moseley2021}, where our data indicate that DR can be neglected compared to destruction by hydrogen collisions, even when the electron fraction is increased to $\chi_\mathrm{e}=10^{-2}$.

Our measurements also indicate that destruction by DR can be enhanced in regions with higher cosmic ray ionization rates or lower than average gas density, where increased electron fractions are found. This is the case in the Galactic center \citep{LePetit2016} or in supernova remnants \citep{Priestley2017}. Figure \ref{fig:plasmarate}(b) contains the boundaries for the DR dominated CH$^{+}$ destruction regime for enhanced electron fractions. We emphasize that these boundaries depend on the behavior of $ \alpha_\mathrm{H} $ and $ \alpha_\mathrm{H_2} $ only for $ T_\mathrm{k}\lesssim\SI{600}{K} $, but not above. Figure \ref{fig:plasmarate}(b) demonstrates that increasing the electron fraction by factors of 6 and 60 compared to quiescent gas in diffuse clouds extends the DR dominated temperature regime to $ T_\mathrm{k}\sim100 $ and $ \SI{500}{K} $ respectively, even for fully molecular clouds. Thus, in environments with increased electron fractions and  $ T_\mathrm{k}\lesssim\SI{500}{K} $, the results of this work are likely to be important for our understanding of the corresponding CH$^{+}$ chemistry. However, for heated regions in diffuse clouds at $T_\mathrm{k}\gtrsim\SI{1000}{K} $, destruction by hydrogen collisions remains the dominant process, even when considering an electron fraction of $ \chi_\mathrm{e}\sim10^{-2} $.

\section{Conclusions} \label{sec:disc}

With this work we have determined the rate coefficient for the most uncertain reaction in the CH$^+$ chemistry of diffuse clouds, namely the DR kinetic temperature rate coefficient. By mimicking diffuse cloud temperatures inside the CSR and conducting a high-resolution electron-ion merged-beams experiment, we have obtained the gas kinetic temperature dependence of the DR rate coefficient for the CH$ ^+(X^1\Sigma^+,v=0,J=0) $ state, involving all temperatures from quiescent to locally heated gas. We further characterized the implications of our experimental results on diffuse cloud chemistry, based on a simple reaction network that includes the experimental rate coefficients for the competing CH$^+$ reactions with atomic and molecular hydrogen. By reducing the uncertainty for the DR rate coefficient from above an order of magnitude to $ \sim\pm20\% $, our findings will improve the reliability of future studies for local heating mechanisms in diffuse clouds based on CH$^{+}$ abundance observations. DR experiments at the CSR with other key molecular ions are in development and could lead to further improvements in our understanding of chemical and physical properties of interstellar clouds.

\section*{Acknowledgements}

Financial support by the Max Planck Society is acknowledged. D.\ P., A.\ K., and D.\ W.\ S.\ were supported in part by the U.S.\ National Science Foundation Division of Astronomical Sciences Astronomy and Astrophysics Grants program under AST-1907188. We thank Zs.\ J.\ Mezei and I.\ F.\ Schneider for providing us their theoretical DR cross sections from \citet{Mezei2019} for comparison of kinetic temperature rate coefficients in this work.
	
\newpage
\appendix 

\section{Electronic, vibrational, and rotational level population evolution} \label{app:rotational}

In order to extract $ J $-specific DR rate coefficients, one needs the storage-time evolution of all contributing CH$^{+}$ electronic, vibrational, and rotational level populations. We have experimentally determined these populations \emph{in situ} for the conditions during our DR measurements (Section \ref{sec:exp}, Appendix \ref{app:collSetup}). 

We injected into the CSR a CH$^{+}$ ion beam with a variety of excited states populated. Most of the electronically excited states are known to radiatively decay with a lifetime of $ \lesssim\SI{10}{\mu s} $ \citep{Huber1979}. The exception to this is the metastable $ a^3{\Pi} $ state, which lacks a spin-allowed transition to the $ X^{1}\Sigma^{+} $ ground state and can only decay via singlet-triplet mixing \citep{Hechtfischer2007}. Due to different coupling constants for individual $ (v,J,e/f) $ states ($ e/f $ being the parity), the $ a^3{\Pi} $ decay is expected to involve a variety of different lifetimes \citep{Hechtfischer2007}. Experimentally, an average lifetime for all contributing states was found to be $ \sim\SI{7}{s} $ \citep{Amitay1996}.

Vibrationally excited states of the $ X^{1}\Sigma^{+} $ and $ a^3{\Pi} $ states radiatively relaxed within the first $ \SI{3}{s} $ of storage in the CSR. For the $ X^{1}\Sigma^{+} $ state, the longest living $ v=1 $ state decays with a lifetime of $ \SI{0.7}{s} $ \citep{Folomeg1987}; while for the $ a^3{\Pi} $ state, after $ \SI{0.5}{s} $ the $ v=1 $ population amounts to less than $ 1\% $ compared to $ v=0 $ \citep{Hechtfischer2007}. Thus, after $ t=\SI{3}{s} $ we expect only $ X^{1}\Sigma^{+} (v=0) $ and $ a^{3}\Pi (v=0) $ in the stored ion beam.

We experimentally characterized the radiative decay of the metastable $ a^3{\Pi} (v=0)$ state by storage-time-dependent CH$^{+}$ DR measurements. Our neutral-fragment detector (Figure \ref{fig:setup}(a)) is capable of imaging the impact positions of neutral C and H fragments. We recorded the distances between coincident C and H fragments from the DR events for a detuning energy of $ E_\mathrm{d}=\SI{0}{eV} $ and over storage times of $ \SI[parse-numbers=false]{3-70}{s} $. The spatial distribution of these distances is due to the energy released in the exothermic DR reaction and is used to determine the kinetic energy of the C and H fragments. This kinetic energy release (KER) is $ >\SI{4}{eV} $ for the $ X^{1}\Sigma^{+}(v=0) $ ground state and $ <\SI{1}{eV} $ for the metastable $ a^{3}\Pi (v=0) $ state (\citealt{Amitay1996}). The very different KER values enable us to follow the relative DR contributions from the $ a^{3}\Pi (v=0) $ state vs.\ storage time and model the metastable storage-time-dependent population $ p_\mathrm{m}(t) $ (see \citealt{Kalosi2022} for further details). We obtained a best match to our data with a two-component decay, involving a short-lived component (lifetime: $ \tau_\mathrm{m}=\SI{10.1\pm1.0}{s} $) and a long-lived component (lifetime $ \gg\SI{100}{s} $), which we consider to have a constant population within the storage times probed, using the equation 
\begin{equation} \label{eq:metModel}
	p_\mathrm{m}(t)=p_{\mathrm{m},t_0}e^{-(t-t_0)/\tau_\mathrm{m}}+p_{\mathrm{m},\infty}.
\end{equation}
We found $ p_{\mathrm{m},t_0}=\SI{0.10\pm0.03}{} $ at $ t_0=\SI{21}{s} $ and $ p_{\mathrm{m},\infty}=\SI{0.025\pm0.007}{} $.

Within the $ X^{1}\Sigma^{+}(v=0) $ state, rotational cooling of $ J\ge3 $ levels was dominated by radiative deexcitation in the $ \sim\SI{20}{K} $ two-component blackbody radiation field of the CSR. Radiative lifetimes of these levels are below $ \SI{5}{s} $ \citep{Kalosi2022}. Inelastic collisions between our electron beam and the CH$^{+}$ ions, with purely matched velocities ($ E_\mathrm{d}=\SI{0}{eV} $) for the initial $ \SI{21}{s} $ of beam storage, accelerate this cooling process via collisional deexcitation. Excitation of levels $ J=0-2 $ into $ J\ge3 $ by inelastic electron collisions is negligible since it requires collision energies $ >\SI{10}{meV} $ \citep{Kalosi2022}. Such collisions are rare for $ E_\mathrm{d}=\SI{0}{eV} $ with a collision energy resolution of $ \sim\SI{2}{meV} $. Thus, at the starting time of our DR measurements ($ t=\SI{21}{s} $) any contribution from $ J\ge3 $ levels can be neglected.

We monitored the rotational population evolution for $ X^{1}\Sigma^{+}(v=0,J=0-2) $ within our DR measurement interval of $ t=\SI[parse-numbers=false]{21-100}{s} $ by probing near-threshold photodissociation of CH$^{+}$ with our OPO laser (Figure \ref{fig:setup}(a)). This process has been studied theoretically and experimentally (\citealt{Hechtfischer2002}; \citealt{OConnor2016}). There is an associated resonant structure for photon wavenumbers of $ \sim\SI{33000}{cm^{-1}} $, lying between the $ \mathrm{C}^{+}(^2P_{1/2})+\mathrm{H}(^2{S}) $ and $ \mathrm{C}^{+}(^2P_{3/2})+\mathrm{H}(^2{S}) $ atomic fine structure levels. These resonances can be assigned to individual initial CH$^{+}(X^{1}\Sigma^{+},v,J)$ levels. By probing the photodissociation cross section at four different photon energies, three of them on top of the strongest $ J=0-2 $ peaks and one for background characterization, we determined the relative rotational populations $ \hat{p}_J(t) [J=0-2] $ within the six storage-time windows probed in the DR measurements (Appendix \ref{app:collSetup}). This method for \emph{in situ} determination of rotational populations was previously applied at the CSR for characterizing radiative cooling by \citet{OConnor2016} and was also utilized for measuring inelastic electron collision cross sections by \citet{Kalosi2022}. Details on the experimental setup and measurement parameters used in the present work can be found in \citet{Kalosi2022}. Here, we applied a lower electron density and a more complex electron collision energy scheme, involving not only collisions at $ E_\mathrm{d}=\SI{0}{eV} $ but also at higher detuning energies, which were required to obtain the DR spectrum. Thus, we have slightly different rotational populations compared to \citet{Kalosi2022}.

\begin{figure}[]
	\plotone{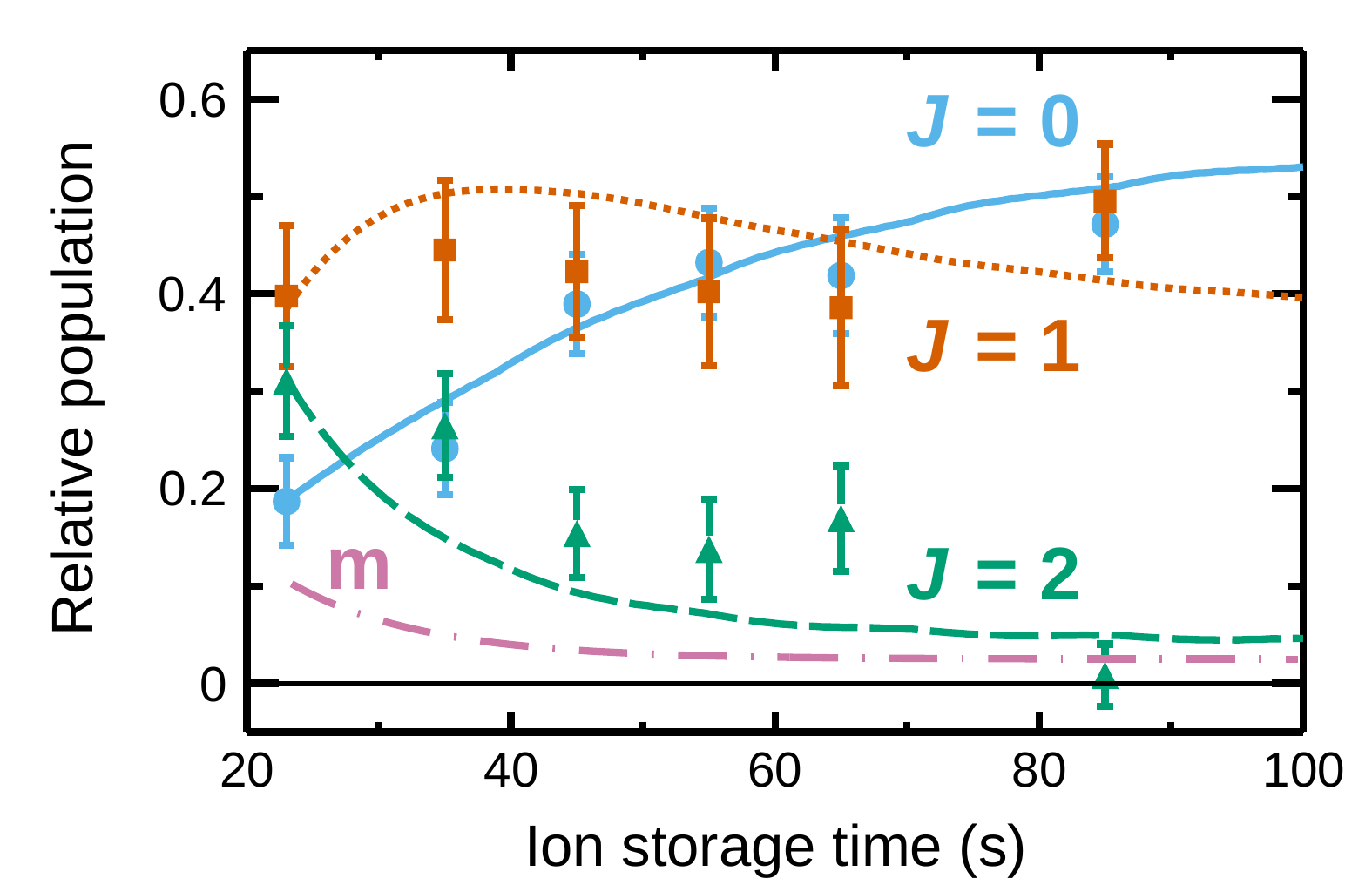}
	\caption{Relative level-population evolution of CH$ ^+ $ ions in the CSR during DR measurements. The markers represent laser-probed rotational populations $ \hat{p}_J $ with statistical one-sigma uncertainties. The lines indicate modeled populations $ p_J $ from experimentally benchmarked radiative rotational cooling \citep{Kalosi2022} and using the $ J $-dependent inelastic electron-collision cross sections for rotational (de\hbox{-})excitation from \citet{Hamilton2015}. Initial populations for the model were fixed to the measured values at $ t=\SI{23}{s} $ (leftmost markers). The metastable $ a^3\Pi $ population evolution ($ p_{\mathrm{m}} $, dash-dotted line) was modeled with Equation (\ref{eq:metModel}). All $ J $ populations were normalized such that $ \sum_{J}\hat{p}_J+p_{\mathrm{m}}=1 $ and $ \sum_{J}p_J+p_{\mathrm{m}}=1$.}
	\label{fig:statePopulations}
\end{figure}

Figure \ref{fig:statePopulations} summarizes the results for the evolution of the metastable $ a^{3}\Pi (v=0) $ and $ X^{1}\Sigma^{+}(v=0,J=0-2) $ populations with storage time in the CSR. In the storage-time interval $ t=\SI[parse-numbers=false]{21-100}{s} $ of our DR measurements, the inferred population of the $ a^{3}\Pi (v=0) $ state is below $ 13\% $. The data for the rotational populations of the  $ X^{1}\Sigma^{+}(v=0)$ state indicate that the $ J=2 $ population decays on a similar time scale as the metastable state but with about a factor of 2 higher population compared to the metastable state. For $ t>\SI{40}{s} $ more than $ 70\% $ of the ions occupy the $ J=0 $ and $ J=1 $ levels.

In order to determine the rotational level populations needed to derive the $ J $-specific DR rate coefficients, we used the combined radiative-cooling and inelastic-collision rotational-cooling models, which yield results in good agreement with the measured rotational populations, as can be seen in Figure \ref{fig:statePopulations}. The combined model is described in detail by \citet{Kalosi2022} and uses the calculated inelastic collision cross sections from \citet{Hamilton2015}. The model was found to be in good agreement with experimental data \citep{Kalosi2022}. Here, we adjusted the model to the inelastic-collision energy scheme used in this work. The resulting modeled populations $ p_J(t) $ are indicated as lines in Figure \ref{fig:statePopulations} and are in agreement with our measured populations $ \hat{p}_J(t) $. The model populations are used in the analysis of the $ J $-specific DR rate coefficients (Appendix \ref{app:stateResolved}). We also used the model to estimate the contribution of $ J\geq3 $ levels for storage times of $ t>\SI{21}{s} $, where collisions at $ E_\mathrm{d}>\SI{0}{eV} $ were applied with an $ \sim18\% $ duty cycle (see Appendix \ref{app:collSetup}). The collisional excitation into $ J\geq3 $ resulted in populations $ p_{J\geq3}<2\% $. We thus neglected the $ J\geq3 $ levels for further analysis.

We note that in the analysis of the $ a^{3}\Pi (v=0) $ population the rotational levels in this electronic state were not considered individually. For the determination of level-specific DR rate coefficients (Section \ref{sec:exp}, Appendix \ref{app:stateResolved}), we justify this simplification by the small relative population of $ a^{3}\Pi (v=0) $ ($ \lesssim 10\% $), which we found has only a minor effect on the rate coefficient analysis for the $ J=0 $ and $ J=1 $ levels.

\section{Electron-ion collision experiments} \label{app:collSetup}

Merged-beams electron-ion collisions in storage rings are frequently used for obtaining collisional rate coefficients of electronic and vibrational ground state ions (e.g., \citealt{Amitay1996}, \citealt{Tanabe1998}, \citealt{Geppert2004}, \citealt{Novotny2013}). However, these experiments were limited to room-temperature environments and, thus, included a broad range of rotational levels of the target molecule. With the commissioning of the CSR, it is now possible to perform merged-beams DR experiments for rotationally cold ions \citep{Novotny2019}. In this work we use this setup for measuring the CH$^{+}$ DR rate coefficient (Figure \ref{fig:setup}(a)).

The CH$^{+}$ ion beam was produced in a Penning ion source and accelerated towards ground potential from an $ \sim\SI{280}{kV} $ high voltage platform \citep{vonHahn2016}. About $ 10^5-10^6 $ ions with a kinetic energy of $ \sim\SI{280}{keV} $ were stored inside the CSR, where they traversed its circumference of $ C_0=\SI{35.12\pm0.05}{m} $ with a revolution frequency of $ f_\mathrm{r}=\SI{57.98\pm0.01}{kHz} $. The storage lifetime of the ions inside the CSR enabled us to perform collision experiments up to $ t=\SI{100}{s} $.

We used the CSR electron cooler for both phase-space cooling of the ion beam and merged-beams DR experiments (Section \ref{sec:exp}). The electron beam was generated from a room-temperature photocathode \citep{Orlov2005} and accelerated to an energy of $ \sim\SI{20}{eV} $ outside of the CSR (Figure \ref{fig:setup}(a)). Over the entire electron cooler section, from the cathode to the collector, the low-energy electron beam was guided by solenoidal magnetic fields, encountering sequentially field strengths of $ \SI{200}{mT} $, $ \SI{\sim100}{mT} $, $ \SI{10}{mT} $ and $ \SI{\sim100}{mT} $. Accordingly, we used a magnetic field gradient with a factor of 20 to adiabatically expand the electron beam \citep{Novotny2013}, resulting in a diameter increase to $ \SI{10.2\pm0.5}{mm} $ and a transverse temperature reduction from $ k_\mathrm{B}T_\perp=\SI{25}{meV} $ (room temperature) to $ k_\mathrm{B}T_\perp=\SI[parse-numbers=false]{2.0^{+1.0}_{-0.5}}{meV} $, which matches previous observations \citep{Kalosi2022}. In the merged-beams interaction region, where the magnetic field strength was $\SI[parse-numbers=false]{10}{mT} $, we controlled the electron beam laboratory-frame energy $ E_\mathrm{e} $ by adjusting the potential of a set of drift tubes (Figure \ref{fig:setup}(a)).

Phase-space cooling of the ion beam was established by matching electron and ion velocities \citep{Poth1990}. The corresponding electron energy $ E_0 $ is given by  
\begin{equation}
	E_0=\frac{1}{2}m_\mathrm{e}f_\mathrm{r}^2C_0^2\sim\SI{11.79}{eV} ,
\end{equation}
where $ m_\mathrm{e} $ is the electron mass. We applied phase-space cooling to the stored CH$^{+}$ ion beam for $ \SI{21}{s} $. During this time window, the Gaussian density profile of the ion beam decreased from an effective FWHM diameter \citep{Kalosi2022} of $ >\SI{8}{mm} $ to an equilibrium of $ \SI{5.2\pm0.5}{mm} $. Thus, in our DR measurements more than $ 90\% $ of the ions were enclosed within the electron beam.

For the collision experiments at $ t=\SI[parse-numbers=false]{21-100}{s} $, we detuned the electron beam laboratory energy $ E_\mathrm{e} $ (with $ E_\mathrm{e}=E_\mathrm{0} $ for phase-space cooling) by adjusting the drift-tube potential. The resulting detuning energy $ E_\mathrm{d} $ as plotted in the merged-beams rate coefficient spectra is derived from
\begin{equation}
	E_\mathrm{d}=\left(\sqrt{E_\mathrm{e}}-\sqrt{E_0}\right)^2 .
\end{equation}
In order not to significantly alter the properties of the phase-space cooled ion beam, we did not dwell at detuned velocities for more than $ \SI{100}{ms} $, but instead switched regularly between matched and detuned velocities. Our standard switching cycle included $ \SI{100}{ms} $ at $ E_\mathrm{d}=\SI{0}{eV} $ (phase-space cooling); $ \SI{25}{ms} $ at a single measurement energy $ E_\mathrm{d}=E_\mathrm{meas} $; and $ \SI{25}{ms} $ at a reference energy $ E_\mathrm{d}=E_\mathrm{ref}\sim\SI{3.4}{eV} $, where the electron-induced rate is small compared to background from residual gas. Between each energy step we implemented an $ \sim\SI{5}{ms} $ waiting time for voltage stabilization. For each value of $ E_\mathrm{d}$, we recorded the number of collisions resulting in neutral particles on our neutral-fragment detector (Figure \ref{fig:setup}(a)), which is based on microchannel plates (MCPs). We also recorded the duration spent at each energy step in order to determine the count rates $ R_\mathrm{meas}(E_\mathrm{d}) $ and $ R_\mathrm{ref} $ for measurement and reference energy, respectively.  The rate at matched velocities ($ R_\mathrm{cool} $) was not measured since when probing $ E_\mathrm{d}>\SI{0.03}{eV} $ with a typically high number of ions in CSR, $ R_\mathrm{cool} $ would  saturate our detector, owing to the high rate coefficient at $ E_\mathrm{d}=\SI{0}{eV} $ (Figure \ref{fig:ratecoeff}). We avoided this saturation by switching off the front MCP voltage in the matched-velocity phases.

Energy-dependent merged-beams DR rate coefficients are calculated from the measured count rates. We first subtract the residual gas background, measured at $ E_\mathrm{ref} $, to determine the electron-induced signal
\begin{equation} \label{eq:rateSubtraction}
	R_\mathrm{e}(E_\mathrm{d})=R_\mathrm{meas}(E_\mathrm{d})-R_\mathrm{ref}.
\end{equation}
The rate coefficient is then given by
\begin{equation} \label{eq:absoluteRate}
	\alpha^\mathrm{mb}(E_\mathrm{d})=\frac{R_\mathrm{e}(E_\mathrm{d})}{\eta_\mathrm{d}(E_\mathrm{d})\zeta n_\mathrm{e}(E_\mathrm{d})N_\mathrm{i}\hat{l}_0/C_0} ,
\end{equation}
where $ \eta_\mathrm{d}(E_\mathrm{d}) $ is the detection efficiency for DR events with our neutral-fragment detector, $ \zeta $ is the fraction of ions enclosed by the electron beam ($ \zeta=1.000^{+0}_{-0.001} $ for the absolute scaling measurement in Appendix \ref{app:absRate}), $ n_\mathrm{e}(E_\mathrm{d}) $ is the electron density, $ N_\mathrm{i} $ is the number of ions in the CSR, and $ \hat{l}_0=\SI{0.79\pm0.01}{m} $ is the effective electron-ion overlap length \citep{Kalosi2022}. Here, the electron density is calculated from the measured electron beam current and radius. 

The detection efficiency $ \eta_\mathrm{d}$ for DR events is determined by two factors. First, the front MCP plate of our detector has a counting efficiency for single particles of $ p_\mathrm{d} =\SI{0.593\pm0.015}{}$. This value was obtained in an independent experiment by evaluating the ratio of single-to-double particle hits resulting from DR reactions of CF$ ^+ $, where nearly all neutrals arrived at the detector surface in coincident pairs (C and F). From the value of $ p_\mathrm{d}$, we calculate an intrinsic DR event detection efficiency, which is the summed probability to detect either one DR fragment or both fragments, as
\begin{equation} 
	\hat{\eta}_\mathrm{d}=2p_\mathrm{d}(1-p_\mathrm{d})+p_\mathrm{d}^2=\SI{0.834\pm0.012}{}.
\end{equation}
Second, taking the geometry into account, the broad spatial distribution of CH$^+$ DR fragments results in a fraction of H atoms arriving outside the detector area. This fraction depends on the contributing DR final channels and, thus, on $ E_\mathrm{d} $. Using the $ p_\mathrm{d} $ value from above, we were able to infer the combined intrinsic and geometric detection efficiency $ \eta_\mathrm{d}(E_\mathrm{d})$ for DR events from measurements of the ratio for single-to-double particle hits $ r_\mathrm{s/d}(E_\mathrm{d}) $ via \citep{Paul2021}
\begin{equation} 
	\eta_\mathrm{d}(E_\mathrm{d})=p_\mathrm{d}+\frac{p_\mathrm{d}-p_\mathrm{d}^2}{p_\mathrm{d}(2+r_\mathrm{s/d}(E_\mathrm{d}))-1}.
\end{equation}
We measured $ r_\mathrm{s/d}(E_\mathrm{d,s}) $ at a few specific detuning energies $ E_\mathrm{d,s} $ and obtained values of $ \eta_\mathrm{d}(E_\mathrm{d,s})\in[0.74,\hat{\eta}_\mathrm{d}] $ with statistical uncertainties at all $E_\mathrm{d,s}$ below $ 0.01 $. When evaluating Equation (\ref{eq:absoluteRate}), we interpolated $ \eta_\mathrm{d}(E_\mathrm{d})$ between the measured $ E_\mathrm{d,s} $. Given the range of possible $ \eta_\mathrm{d} $ values, we expect the systematic uncertainty of this interpolation to be $ <5\% $ and, thus, to be negligible compared to our absolute scaling uncertainty (Appendix \ref{app:absRate}). See \citet{Paul2021} for further details on the DR event detection efficiency analysis.

The storage-time dependence of the merged-beams rate coefficient $ \alpha^\mathrm{mb}(E_\mathrm{d},t) $ is found using Equation (\ref{eq:absoluteRate}) for certain storage-time windows, where we consider the storage-time dependences of $ R_\mathrm{e}(t) $ and $ N_\mathrm{i}(t) $. Since we did not continuously measure the ion number, we use the detected background rate $ R_\mathrm{b}(t) $ from collisions of CH$^+$ with residual-gas particles as a proxy for the ion number. This rate is inferred from $ R_\mathrm{b}(t)=R_\mathrm{ref}(t)-R_\mathrm{dark} $, where $ R_\mathrm{dark}\sim\SI{50}{Hz} $ is the experimentally determined dark-count rate of our detector. We can then calculate a relative merged-beams rate coefficient
\begin{equation} \label{eq:relativeRate}
	\alpha^\mathrm{mb}_\mathrm{rel}(E_\mathrm{d},t)=\frac{1}{\eta_\mathrm{d}(E_\mathrm{d})\zeta n_\mathrm{e}(E_\mathrm{d})\hat{l}_0/C_0}\frac{R_\mathrm{e}(E_\mathrm{d},t)}{R_\mathrm{b}(t)},
\end{equation}
for each storage-time window within a given measurement set. We evaluate Equation (\ref{eq:relativeRate}) for each measurement set separately, and combine different sets by scaling the results so that they match at commonly measured ranges for $ E_\mathrm{d} $ and $ t $. Finally, we determine the dependence $ N_\mathrm{i}(R_\mathrm{b}) $ for a single set to obtain the absolute rate coefficient in Equation (\ref{eq:absoluteRate}) (see Appendix \ref{app:absRate}).

The energy resolution in our merged-beams rate coefficient spectra, as shown in Figure \ref{fig:ratecoeff}, is dominated by the velocity spread within the electron beam, which exceeds that of the phase-space cooled CH$^+$ ion beam. The corresponding probability distribution $ f_\mathrm{mb}(E;E_\mathrm{d}) $ of collision energies $ E $ for a single detuning energy $ E_\mathrm{d} $ is influenced by several factors (see Figure \ref{fig:setup}(a)): the angles between the electron and ion beams when merging and de-merging the two beams; the electron energy transition from outside to inside the drift tubes; the transverse temperature of the electron beam of $ k_\mathrm{B}T_\perp=\SI[parse-numbers=false]{2.0^{+1.0}_{-0.5}}{meV} $; and the longitudinal temperature of the electron beam $ k_\mathrm{B}T_\parallel(E_\mathrm{e}) $, which itself varies with the laboratory-frame energy of the electron beam ($ E_\mathrm{e} $), i.e., over the full overlap section. This variation is due to the kinematic compression effect when accelerating a particle beam, and due to relaxation effects within the beam that minimize the potential energy through Coulomb interactions \citep{Paul2021}. 

To account for all these effects, we have developed an analytic model, starting from electron emission by the photocathode, and accounting for subsequent acceleration and deceleration steps, to calculate $ k_\mathrm{B}T_\parallel(E_\mathrm{e})$. Furthermore, we use a Monte-Carlo simulation in order to model the full collision geometry. An example for a resulting energy distribution function $ f_\mathrm{mb}(E;E_\mathrm{d}) $ is found in \cite{Kalosi2022}. In Appendix \ref{app:deconvolution} we use $ f_\mathrm{mb}(E;E_\mathrm{d}) $ to derive the DR cross section from our merged-beams rate coefficient. The CSR energy resolutions displayed in Figures \ref{fig:ratecoeff} and \ref{fig:ratecoeffmeas} correspond to the FWHM of $ f_\mathrm{mb}(E;E_\mathrm{d}) $.

\section{Absolute scaling of the merged-beams rate coefficient} \label{app:absRate}

Absolute scaling of the merged-beams DR rate coefficient requires knowledge of the number of ions stored inside the CSR. Since we could not measure the ion number at all storage times, we used the storage-time dependence of the background rate $ R_\mathrm{b}(t)$ as a proxy for the ion number evolution (Appendix \ref{app:collSetup}). The proportionality factor 
\begin{equation} \label{eq:propfactor}
	S_\mathrm{b}=R_\mathrm{b}(t)/N_\mathrm{i}(t)
\end{equation}
was evaluated in two separate measurements. First, we calibrated the voltage induced by the ion beam on a capacitive current pickup in the CSR (PU-C; \citealt{vonHahn2016}) to the ion current $ I $. Second, we determined the averaged ion number $ \langle N_\mathrm{i}\rangle $ from $ I $ in a small storage-time window $ t=\SI[parse-numbers=false]{21-21.5}{s} $, together with the averaged residual gas background rate $ \langle R_\mathrm{b}\rangle $.

For the ion current calibration of the PU-C, we compared the induced voltage to a direct ion current measurement using a Faraday cup, which is located in front of the CSR. A fast-switched ($ <\SI{50}{ns} $) electrostatic deflector was used to create an $ \sim\SI{20}{\mu s} $ long square-wave pulse of CH$^{+}$ ions from the ion source. The square wave was alternately dumped into the Faraday cup to determine the ion current and injected into the CSR to detect the corresponding induced voltage on the PU-C in the first revolution. We verified that the Faraday cup collection efficiency was essentially unity ($ \pm2\% $). By moving the ion beam across the Faraday cup aperture, we obtained a flat plateau in the ion current. We prevented the loss of secondary electrons by applying a negative voltage of $\SI{-250}{V}$ to an aperture placed directly in front of the Faraday cup, where further increase of the voltage did not change the measured ion current. Additionally, we verified that there is an efficient transport into the CSR by reducing the beam emittance with a set of $ \SI{4.5}{mm} $ apertures in the injection beamline and repeating the calibration. The lower emittance beam is expected to be injected and transported to the pickup similarly or more efficiently than our usual ion beams. We obtained calibration factors agreeing within $1\% $  for the two beams. A similar agreement was obtained when reducing the ion current by a factor of $ \sim10 $, which indicates sufficient linearity for the calibration. We estimate the final uncertainty in the ion current calibration of the PU-C to be $ 10\% $, mainly due to the stability of the ion source between square wave pulses.

With the calibrated PU-C, we were able to determine $ S_\mathrm{b} $ from Equation (\ref{eq:propfactor}) in a separate measurement. For this, the ion beam was stored inside the CSR and bunched by applying an AC voltage with a frequency of $ \sim\SI{400}{kHz} $ with the CSR rf-bunching system \citep{vonHahn2016}, which is located in the left section of the CSR in Figure \ref{fig:setup}(a). This voltage caused the stored ions to split into seven distinct bunches. When passing the PU-C, each bunch induced a mirror charge. The corresponding time-dependent voltage was converted into an ion current  $ I(t) $. The absolute ion number in the storage-time interval $ t=\SI[parse-numbers=false]{21-21.5}{s}$ is then given by
\begin{equation} 
	\langle N_\mathrm{i}\rangle =\frac{1}{Zen_\mathrm{r}}\int_{\SI{21}{s}}^{\SI{21.5}{s}}I(t)\mathrm{d}t,
\end{equation}
where $Z=1$ is the charge state of CH$^+$, $ e $ is the elementary charge, and $ n_\mathrm{r}\sim30000 $ is the number of revolutions the ions made in the CSR during the measurement time. Finally, we calculate the proportionality factor between ion number and residual-gas-induced detector count rate using Equation (\ref{eq:propfactor}).

The absolute CH$^{+}$ DR rate coefficient for energies of $ E_\mathrm{d}=\SI[parse-numbers=false]{0.06-0.29}{eV} $ was determined simultaneously with the bunched-beam ion-number measurement and extended to a storage-time interval of $ t=\SI[parse-numbers=false]{21-100}{s} $. After the bunched-beam ion-number measurement, we removed the AC voltage from the rf system and obtained a continuous ion beam at $ t>\SI{21.5}{s} $, similar to all other DR measurement sets. The rate coefficient $ \alpha^\mathrm{mb}(E_\mathrm{d}=\SI[parse-numbers=false]{0.06-0.29}{eV},  t=\SI[parse-numbers=false]{21-100}{s}) $ was then calculated from Equation (\ref{eq:absoluteRate}), using $ N_\mathrm{i}(t)=R_\mathrm{b}(t)/S_\mathrm{b} $. 

Next, we put the relative rate coefficient for all measurement sets (Appendix \ref{app:collSetup}) onto an absolute scale by matching $  \alpha^\mathrm{mb}_\mathrm{rel}(E_\mathrm{d}=\SI[parse-numbers=false]{0.06-0.29}{eV},  t=\SI[parse-numbers=false]{21-100}{s})$ to $  \alpha^\mathrm{mb}(E_\mathrm{d}=\SI[parse-numbers=false]{0.06-0.29}{eV},  t=\SI[parse-numbers=false]{21-100}{s})$. The total uncertainty of the absolute scaling we estimate to be $ 13\% $, where the main contributions stem from the ion current calibration mentioned above ($10\%$) and the electron density uncertainty ($8\%$).

A small additive correction of $ \delta\alpha^\mathrm{mb}=\SI{6.1e-9}{cm^3\,s^{-1}} $ is applied to the absolute scaled rate coefficient due to the fact that $ R_\mathrm{ref} $ (see Appendix \ref{app:collSetup}) was due to both residual-gas-induced events and a small contribution from DR. We determined $ \delta\alpha^\mathrm{mb}$ in a separate experiment by measuring the absolute rate coefficient at $ E_\mathrm{ref} $ using 
\begin{equation} 
	R_\mathrm{e}(E_\mathrm{ref})=R_\mathrm{ref}-R_\mathrm{res}.
\end{equation}
instead of Equation (\ref{eq:rateSubtraction}). Here, $ R_\mathrm{res} $ stands for the pure residual-gas-induced rate, which we determined by switching off the electron beam in a fast manner, similar to the fast energy adjustment described in Appendix \ref{app:collSetup}. The correction $ \delta\alpha^\mathrm{mb} $ is due to DR reactions in the electron-ion beam-overlap section outside of the drift tubes in Figure \ref{fig:setup}(a). Thus, it occurs at all $ E_\mathrm{d} $ in the merged-beams rate coefficient (Figure \ref{fig:ratecoeff}). The collisions outside the drift-tube section take place at collision energies of $E\sim\SI{1.2}{eV} $, located on the flank of the DR rate coefficient peak in Figure \ref{fig:ratecoeff}, and are included in the collision energy distribution $ f_\mathrm{mb}(E;E_\mathrm{d}) $. The offset $ \delta\alpha^\mathrm{mb} $ is accounted for when calculating the DR cross section in Appendix \ref{app:deconvolution}.

\section{Evaluation of level-resolved merged-beams rate coefficients}  \label{app:stateResolved}

\begin{figure*}[]
	\epsscale{0.7}
	\plotone{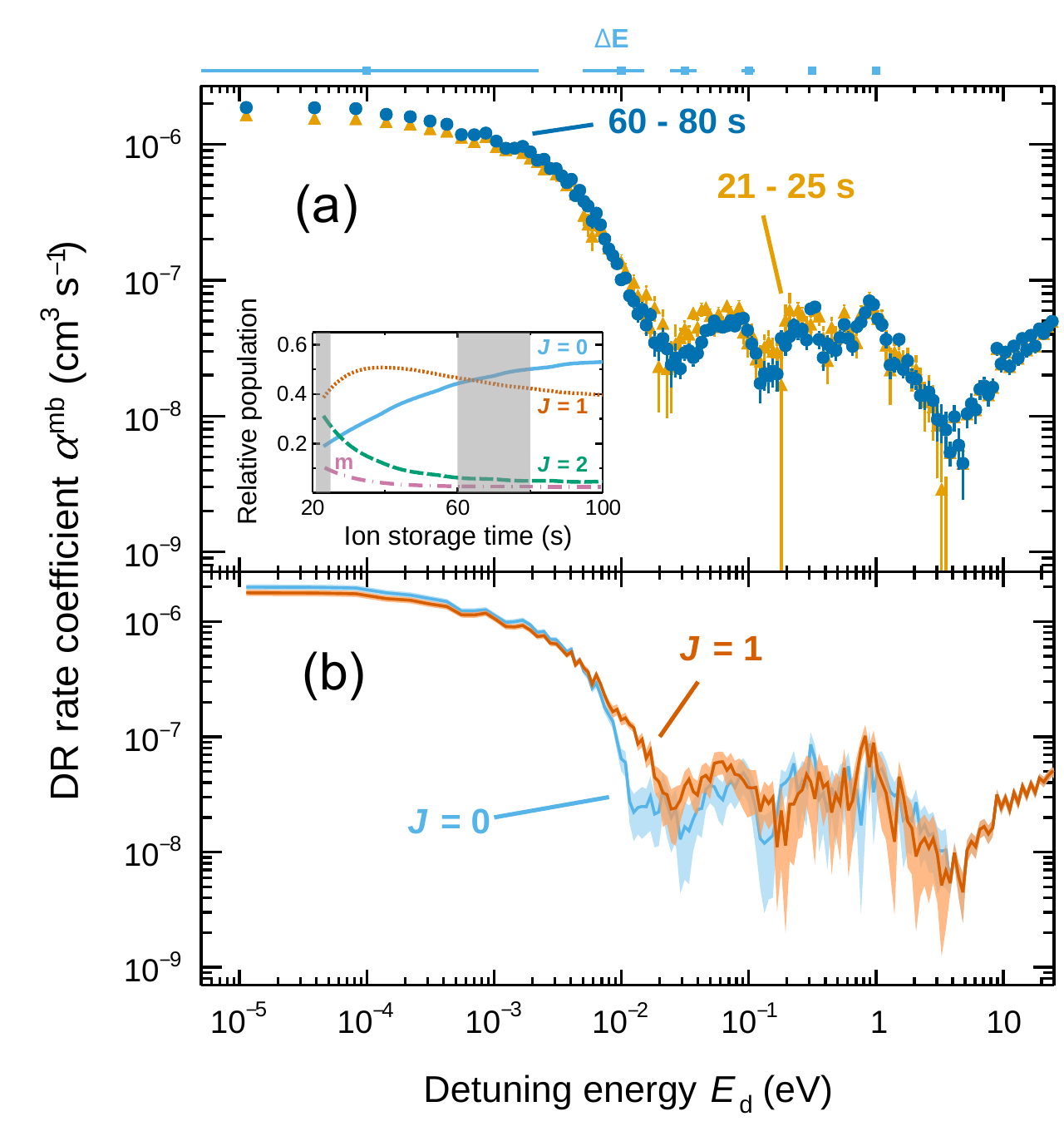}
	\caption{(a) Merged-beams DR rate coefficient of CH$^{+}$ for two different ion-beam storage-time windows in the CSR. Relative rotational populations in the CSR (see Figure \ref{fig:statePopulations}) are shown in the inset. The gray areas highlight the storage times for the measurements shown. (b) Comparison of the $ J=0 $ and $ J=1 $ rate coefficients calculated using a MCMC method (see text). In both panels, statistical one-sigma uncertainties are indicated by the error bars and bands, while the absolute scaling uncertainty is $ \pm13 \% $. The data behind panel (b) are available.}
	\label{fig:ratecoeffmeas}
\end{figure*}

Independent control over the internal ion-beam levels and electron collision energies in the CSR enables us to determine rotational level-resolved DR merged-beams rate coefficients, as was first demonstrated by \cite{Novotny2019}. Here, we use a similar method to evaluate the CH$^+$ DR merged-beams rate coefficient of the $ X^{1}\Sigma^{+}(v=0,J=0,1) $ levels.

The evaluation of $ J $-specific DR rate coefficients relies on storage-time-dependent DR measurements and knowledge of all internal level populations in each storage-time window. Here, we have binned the absolute DR rate coefficient ($ \alpha^\mathrm{mb}(E_\mathrm{d},t) $) into six storage-time intervals
\begin{eqnarray} 
	t[s]\in\{&21-25,&25-30,30-40,\nonumber\\
	&40-60,&60-80,80-100\},
\end{eqnarray}
which are denoted as $ t_k $ $(k\in\{1,...,6\})$  in the following. Two examples for the rate coefficients in different time windows are shown in Figure \ref{fig:ratecoeffmeas}(a). The corresponding rotational ($ p_J(t_k)[J=0-2] $) and metastable ($ p_\mathrm{m}(t_k) $) populations within these storage-time windows are taken from the modeled lines in Figure \ref{fig:statePopulations} and are highlighted in the inset of Figure \ref{fig:ratecoeffmeas}(a). Here, we used the averaged values within each storage time window $ t_k $ for further analysis. The relation between level-specific rate coefficients $ \alpha^\mathrm{mb}_{J/\mathrm{m}}(E_\mathrm{d}) $ and the storage-time-dependent ones is given by the linear equation system
\begin{equation} 
	\alpha^\mathrm{mb}(E_\mathrm{d},t_k)=\sum_{J=0}^{2} \alpha^\mathrm{mb}_J(E_\mathrm{d})p_J(t_k)+\alpha^\mathrm{mb}_\mathrm{m}(E_\mathrm{d})p_\mathrm{m}(t_k).
\end{equation}

Using our six storage-time-dependent results, one obtains an overdetermined system for evaluating the four level-specific rate coefficients. In principle, such a system could be solved by least-squares minimization. However, a least-squares approach is not suited for including the boundary conditions $ \alpha^\mathrm{mb}_{J/\mathrm{m}}(E_\mathrm{d})\ge0 $, which are important for the $ J=2 $ and metastable levels, which both have a small but non-zero population. As an alternative, we use a Markow Chain Monte Carlo (MCMC; \citealt{Hogg2018}) method for the statistical analysis, which is based on the Bayesian inference principle and is capable of embedding the required boundary conditions. The result of the MCMC algorithm is the posterior probability distribution $ p(\alpha^\mathrm{mb}_{J/\mathrm{m}}|D) $ for the rate coefficients based on the observed data $ D $. The posterior is calculated from
\begin{equation} 
	p(\alpha^\mathrm{mb}_{J/\mathrm{m}}|D)=Ze^{-\chi_\nu^2/2}p(\alpha^\mathrm{mb}_{J/\mathrm{m}}),
\end{equation}
where $ Z $ is a normalization constant, $ e^{-\chi_\nu^2/2} $ is the log-likelihood function, and $ p(\alpha^\mathrm{mb}_{J/\mathrm{m}}) $ is the prior probability distribution for the rate coefficients. The reduced chi-squared $ \chi_\nu^2 $ for $ \nu $ degrees of freedom enters the log-likelihood function and is calculated from the rate coefficient data $ \alpha^\mathrm{mb}(t_k) $ as
\begin{equation} \label{eq:chi2}
	\chi_\nu^2=\frac{1}{\nu}\sum_{k=1}^{6}\left(\frac{\alpha^\mathrm{mb}(t_k)-\sum_{J/\mathrm{m}} \alpha^\mathrm{mb}_{J/\mathrm{m}}p_{J/\mathrm{m}}(t_k)}{\Delta \alpha^\mathrm{mb}(t_k) }\right)^2,
\end{equation}
where $ \Delta \alpha^\mathrm{mb}(t_k) $ are the statistical one-sigma uncertainties of our measured $ \alpha^\mathrm{mb}(t_k) $. We account for the detuning energy dependence of the rate coefficients $ \alpha^\mathrm{mb}(E_\mathrm{d},t_k) $ by running a separate MCMC analysis for each measured detuning energy. MCMC algorithms are designed for sampling the log-likelihood function efficiently and are thus well suited for multidimensional parameter spaces.

For this analysis we used the MCMC python package \textsc{emcee} \citep{Foreman-Mackey2013}. We applied the above mentioned boundary conditions and set the initial prior to a uniform probability distribution as
\begin{equation} 
	p(\alpha^\mathrm{mb}_{J/\mathrm{m}})=\left\{\begin{array}{ll}
		Z_0  \quad\qquad & 0\leq\alpha^\mathrm{mb}_{J/\mathrm{m}}\leq\SI[parse-numbers=false]{10^{-5}}{cm^3\,s^{-1}}, \\ 
		0 \quad\qquad &\text{otherwise} ,\\
	\end{array} 
	\right.
\end{equation}
where $ Z_0 $ is a normalization constant. Using our measured $ \alpha^\mathrm{mb}(E_\mathrm{d},t_k) $ led to a high statistical scatter of the results. We avoided this by constructing a moving average of our $ \alpha^\mathrm{mb}(E_\mathrm{d},t_k) $ and $ \Delta\alpha^\mathrm{mb}(E_\mathrm{d},t_k) $ data over the experimental FWHM energy resolution (see bars on top of Figure \ref{fig:ratecoeffmeas}) and using these average quantities in Equation (\ref{eq:chi2}). We justify this averaging procedure for the MCMC input data by the fact that any level-specific features in the data must be visible over a range of detuning energies larger than our energy resolution. 

The MCMC results are the probability distributions $ p(\alpha^\mathrm{mb}_{J/\mathrm{m}}|D) $ for the individual rate coefficients. For $ \alpha^\mathrm{mb}_{J=0,1} $ we obtain sufficiently narrow distributions and are able to evaluate their most probable values and one-sigma uncertainties $\Delta \alpha^\mathrm{mb}_{J=0,1} $. These results are presented in Figure \ref{fig:ratecoeffmeas}(b) as lines and error bands, respectively. We note that the MCMC analysis was only applied to the rate coefficient data up to $ E_\mathrm{d}=\SI{3.6}{eV} $, corresponding to the rate coefficient minimum before the onset of the direct DR/DE peak in Figure \ref{fig:ratecoeff}. At higher energies we do not expect any $ J $-dependent features due to the non-resonant character of the DR and DE process (Section \ref{sec:res}). Hence, at these energies, we set  $ \alpha^\mathrm{mb}_{J=0,1}(E_\mathrm{d})=\alpha^{\mathrm{mb}}(E_\mathrm{d},\SI[parse-numbers=false]{21-100}{s}) $ and $ \Delta\alpha^\mathrm{mb}_{J=0,1}(E_\mathrm{d})=\Delta\alpha^{\mathrm{mb}}(E_\mathrm{d},\SI[parse-numbers=false]{21-100}{s}) $, leading to the narrowed error band above $ E_\mathrm{d}=\SI{3.6}{eV} $. For the $ J=2 $ and metastable levels, the MCMC analysis does not yield significant results. Due to their low population, a broad range of values is accepted for $ \alpha^\mathrm{mb}_{J=2} $ and $ \alpha^\mathrm{mb}_\mathrm{m} $. 

\section{Obtaining kinetic temperature rate coefficients from merged-beams rate coefficient data} \label{app:deconvolution}

The conversion of our CH$^{+}$ $ (J=0,1) $ DR merged-beams rate coefficients $ \alpha^\mathrm{mb}_{J}(E_\mathrm{d}) $ into kinetic temperature rate coefficients $ \alpha^\mathrm{k}_{J}(T_\mathrm{k}) $, as shown in Figure \ref{fig:plasmarate}(a), is achieved in a two-step process. First, we deconvolve\footnote[4]{The terms convolution and deconvolution in this section are not used in a rigorous mathematical meaning. They refer to an integral transformation of the cross section to a rate coefficient (convolution; Equation (\ref{eq:convolutionShort})) and its inverse (deconvolution). Both transformations are described in detail by \citet{Novotny2013}.} the merged-beams rate coefficients into cross sections $ \sigma_J(E) $, using our merged-beams collision energy resolution function $ f_\mathrm{mb}(E;E_\mathrm{d})$ (Appendix \ref{app:collSetup}). Second, we convolve the resulting cross sections with a Maxwell-Boltzmann distribution to obtain the kinetic rate coefficients.

In the deconvolution process we make use of the relation between cross sections and merged-beams rate coefficients (Equation (\ref{eq:convolutionShort})) that can be explicitly expressed as
\begin{equation} \label{eq:convolutionLong}
	\alpha^\mathrm{mb}_{J}(E_\mathrm{d})=\frac{l_0}{\hat{l}_0}\int_{0}^{\infty} \sigma_J(E) \sqrt{\frac{2E}{m_\mathrm{e}}}  f_\mathrm{mb}(E;E_\mathrm{d}) \mathrm{d}E,
\end{equation}
where $ l_0=\SI{1.123\pm0.014}{m} $ is the total geometrical overlap length of electrons and ions. Using the deconvolution method of \citet{Novotny2013}, we can generate a cross section histogram $ \sigma_i $ that is binned in the energy domain. This method relies on fitting the contents of the individual cross section bins such that the discrete integral version of Equation (\ref{eq:convolutionLong}) reproduces the merged-beams rate coefficient data best. 

Here, we have improved the procedure of \citet{Novotny2013} in two ways. First, we use a more detailed collision energy distribution $ f_\mathrm{mb}(E;E_\mathrm{d}) $ that involves not only the geometry of our drift-tube collision setup (Figure \ref{fig:setup}(a)) but also the variation of the electron beam longitudinal temperature $k_\mathrm{B}T_\parallel(E_\mathrm{e}) $ (Appendix \ref{app:collSetup}). Second, we model the energy dependence of the cross section for $ E<\SI{0.5}{meV} $, where our energy resolution does not allow us to experimentally determine any dependence, as
\begin{equation} \label{eq:crossSectionLowEnergy}
	\sigma_0(E)=\sigma_{0,\mathrm{c}}/E,
\end{equation}
instead of assuming a constant value. Here, $ \sigma_{0,\mathrm{c}} $ is fitted in the deconvolution process. Equation (\ref{eq:crossSectionLowEnergy}) accounts for the possibility of direct DR at low collision energies and includes the diverging nature of the DR cross section towards $E=0$ \citep{Mitchell1990}. We note that  assuming instead a constant cross section for $ E<\SI{0.5}{meV} $ does not influence the resulting kinetic rate coefficient significantly.

The kinetic rate coefficients $ \alpha^\mathrm{k}_{J}(T_\mathrm{k}) $ are calculated from a convolution of the DR cross sections with a Maxwell-Boltzmann distribution of temperature $ T_\mathrm{k} $ as 
\begin{equation} \label{eq:plasmaRateConv}
	\alpha^\mathrm{k}_{J}(T_\mathrm{k})=\int_{0}^{\infty} \sigma_J(E) \sqrt{\frac{2E}{m_\mathrm{e}}} \sqrt{\frac{4E}{\pi(k_\mathrm{B}T_\mathrm{k})^3}}e^{-E/k_\mathrm{B}T_\mathrm{k}} \mathrm{d}E.
\end{equation}
Due to the discrete character of our cross sections, we evaluate Equation (\ref{eq:plasmaRateConv}) as a sum instead of an integral to generate the CSR kinetic rate coefficient curves (e.g., Figure \ref{fig:plasmarate}(a)).

The kinetic rate coefficients are effected by several uncertainties. Statistical uncertainties of the merged-beams rate coefficients (see error band in Figure \ref{fig:ratecoeff}) were propagated through the deconvolution and convolution procedure into $ \alpha^\mathrm{k}_{J}(T_\mathrm{k}) $ and are negligible compared to the systematic uncertainties. We account for the systematic uncertainties discussed below and added them in quadrature to generate the kinetic rate coefficient error bands (see Figure \ref{fig:plasmarate}(a)). The absolute scaling uncertainty of $ \alpha^\mathrm{mb}_{J} $ ($ \pm13\% $; Appendix \ref{app:absRate}) propagates directly into $ \alpha^\mathrm{k}_{J}(T_\mathrm{k}) $ and is independent of temperature. It dominates the total uncertainty for $ \SI{650}{K}<T_\mathrm{k}<\SI{20000}{K} $. We also ran our deconvolution and convolution procedures for different longitudinal ($ k_\mathrm{B}T_\parallel $) and transverse temperatures ($ k_\mathrm{B}T_\perp $) of our electron beam. Here, the influence of $ k_\mathrm{B}T_\parallel$ turns out to be negligible. For $ k_\mathrm{B}T_\perp $, we vary it between extremes of $ \SI{1.5}{meV} $ and $ \SI{3}{meV} $, where $ k_\mathrm{B}T_\perp=\SI{1.5}{meV} $ results in a common decrease of all kinetic rate coefficient values and $ k_\mathrm{B}T_\perp=\SI{3}{meV} $ results in a common increase. The corresponding uncertainties are the main contribution for $ T_\mathrm{k}<\SI{650}{K} $ and result in a broadening of the error band towards lower temperatures. Finally, we evaluate the influence of the merged-beams rate coefficients at high energies $\alpha^\mathrm{mb}_{J}(E_\mathrm{d}>\SI{3.6}{eV}) $, where we cannot distinguish between DR and DE (see Section \ref{sec:res}). To account for any possible DE contribution, we determine the kinetic rate coefficient for $ \alpha^\mathrm{mb}_{J}(E_\mathrm{d}>\SI{3.6}{eV}) $ scaled to $ 0 $. The corresponding uncertainty becomes dominant for $ T_\mathrm{k}>\SI{20000}{K} $, leading to the broadening of the lower uncertainty limit with increasing temperature in Figure \ref{fig:plasmarate}(a).

We also note that for constructing the theoretical kinetic rate coefficient curve in Figure \ref{fig:plasmarate}(a) we used the calculated cross section data from \cite{Mezei2019}, which were provided to us by the authors. We convolved their cross section into a kinetic rate coefficient via Equation (\ref{eq:plasmaRateConv}), similar to the treatment of our data. The result is shown as dotted line in Figure \ref{fig:plasmarate}(a).

\section{Analytic representations of the $J=0,1$ DR kinetic temperature rate coefficients} \label{app:plasmaRateResults}

The results for the CH$^+(J=0,1)$ DR kinetic temperature rate coefficients can be incorporated into astrochemical databases or used as input for chemical network simulations. For this purpose, we provide two different analytic representations for the rate coefficient and its one-sigma error bands  (Figure \ref{fig:plasmarate}(a)).

First, we model our data, as well as the upper and lower one-sigma error boundaries, with the fit function proposed by \cite{Novotny2013},
\begin{eqnarray} \label{eq:plasmaRateFunction2}
	\alpha^{\mathrm{k}}_J(T_{\mathrm{k}})[\mathrm{cm^3\,s^{-1}}]&=&
	A\left(\frac{300}{T_{\mathrm{k}}[\mathrm{K}]}\right)^n\\ 
	&+&T_{\mathrm{k}}[\mathrm{K}]^{-1.5}\sum_{r=1}^{4}c_r\exp\left(-\frac{T_r}{T_{\mathrm{k}}[\mathrm{K}] }\right), \nonumber
\end{eqnarray}
where $ A $, $ n $, $ c_r $, and $ T_r $ are the fit parameters. Their resulting values are presented in Tables \ref{tab:plasmaRateOldA} and \ref{tab:plasmaRateOldAJ1} for the $ J=0 $ and $J=1$ levels, respectively. The relative deviations of the listed fit functions from the rate coefficients and their lower and upper error bounds are below $2.3\%$ at all temperatures. The representation in Equation (\ref{eq:plasmaRateFunction2}) explicitly considers features (i.e., broad peaks or dips resulting from multiple resonances in the cross section) in the rate coefficient by the summation term, where each $ r $ stands for one of the $ N_r=4 $ features. This enables an appropriate modeling of the data with a low number of fit parameters ($2(N_r+1) $). Additionally, Equation (\ref{eq:plasmaRateFunction2}) is a continuously differentiable function, which is generally desired for chemical modeling purposes.

\begin{deluxetable}{cccc} \label{tab:plasmaRateOldA}
	\tabletypesize{\scriptsize}
	\tablecaption{Fit parameters for the CH$^{+}$($ J=0 $) kinetic temperature rate coefficient and its lower and upper error bound in Figure \ref{fig:plasmarate}(a), using Equation (\ref{eq:plasmaRateFunction2}) as the fit function.}
	\tablewidth{0pt}
	\tablehead{
		\colhead{Parameter} & \colhead{Rate coefficient} & \colhead{Lower error limit} & \colhead{Upper error limit}
	}
	\startdata
	$A$       & $\SI{2.43e-7}{}$       & $\SI{2.07e-7}{}$    & $\SI{2.94e-7}{}$        \\ 
	$n$       & $\SI{0.740}{}$          & $\SI{0.719}{}$      & $\SI{0.761}{}$          \\ 
	$c_1$     & $\SI{0.642}{}$       & $\SI{-0.327}{}$   & $\SI{0.660}{}$       \\ 
	$c_2$     & $\SI{2.36e-2}{}$        & $\SI{1.73e-2}{}$   & $\SI{2.85e-2}{}$        \\ 
	$c_3$     & $\SI{-2.58e-3}{}$       & $\SI{-2.48e-3}{}$   & $\SI{-3.09e-3}{}$       \\ 
	$c_4$     & $\SI{-1.13e-3}{}$        & $\SI{-1.10e-3}{}$    & $\SI{-1.29e-3}{}$        \\ 
	$T_1$     & $\SI{112000}{}$         & $\SI{151000}{}$      & $\SI{107000}{}$         \\ 
	$T_2$     & $\SI{12000}{}$          & $\SI{12400}{}$       & $\SI{11800}{}$          \\ 
	$T_3$     & $\SI{941}{}$           & $\SI{1050}{}$       & $\SI{919}{}$           \\ 
	$T_4$     & $\SI{220}{}$            & $\SI{234}{}$        & $\SI{215}{}$            \\
	\enddata
\end{deluxetable}

\begin{deluxetable}{cccc} \label{tab:plasmaRateOldAJ1}
	\tabletypesize{\scriptsize}
	\tablecaption{Same as Table \ref{tab:plasmaRateOldA} but for CH$^{+}$($ J=1 $).}
	\tablewidth{0pt}
	\tablehead{
		\colhead{Parameter} & \colhead{Rate coefficient} & \colhead{Lower error limit} & \colhead{Upper error limit}
	}
	\startdata
	$A$       & $\SI{2.22e-7}{}$       & $\SI{2.51e-7}{}$    & $\SI{2.62e-7}{}$        \\ 
	$n$       & $\SI{0.733}{}$          & $\SI{0.613}{}$      & $\SI{0.762}{}$          \\ 
	$c_1$     & $\SI{1.02}{}$       & $\SI{-0.183}{}$   & $\SI{1.07}{}$       \\ 
	$c_2$     & $\SI{3.41e-2}{}$        & $\SI{-7.36e-3}{}$   & $\SI{4.20e-2}{}$        \\ 
	$c_3$     & $\SI{-2.56e-3}{}$       & $\SI{-2.36e-3}{}$   & $\SI{-2.65e-3}{}$       \\ 
	$c_4$     & $\SI{-1.02e-3}{}$        & $\SI{-6.82e-4}{}$    & $\SI{-1.10e-3}{}$        \\ 
	$T_1$     & $\SI{108000}{}$         & $\SI{65600}{}$      & $\SI{104000}{}$         \\ 
	$T_2$     & $\SI{11900}{}$          & $\SI{2300}{}$       & $\SI{11700}{}$          \\ 
	$T_3$     & $\SI{1260}{}$           & $\SI{587}{}$       & $\SI{1160}{}$           \\ 
	$T_4$     & $\SI{264}{}$            & $\SI{161}{}$        & $\SI{255}{}$            \\
	\enddata
\end{deluxetable}

\begin{deluxetable*}{cccccc} \label{tab:plasmaRateKIDA}
	\tabletypesize{\small}
	\tablewidth{\textwidth}
	\tablecaption{Fit parameters for the CH$^{+}$($ J=0 $) kinetic temperature rate coefficient and its relative uncertainty in Figure \ref{fig:plasmarate}(a), using Equations (\ref{eq:plasmaRateFitKIDA}) and (\ref{eq:plasmaRateUncertaintyFitKIDA}) respectively, as fit functions for different temperature ranges.}
	\tablehead{
		\colhead{Parameter} & \multicolumn{5}{c}{Temperature range} \\  & \colhead{$ \SI[parse-numbers=false]{10-69.6}{K} $} & \colhead{$ \SI[parse-numbers=false]{69.6-469}{K} $} & \colhead{$ \SI[parse-numbers=false]{469-3920}{K} $} & \colhead{$ \SI[parse-numbers=false]{3920-19900}{K} $} & \colhead{$ \SI[parse-numbers=false]{19900-40000}{K} $}
	}
	\startdata
	$A$       & $\SI{1.77e-7}{}$       & $\SI{1.16e-7}{}$    & $\SI{2.03e-8}{}$  & $\SI{3.08e-7}{}$ & $\SI{1.70e-10}{}$      \\ 
	$\beta$       & $\SI{-0.915}{}$       & $\SI{-1.13}{}$    & $\SI{7.28e-2}{}$  & $\SI{-0.689}{}$  &  $\SI{0.756}{}$   \\
	$\gamma$       & $\SI{2.94}{}$       & $\SI{-4.59}{}$    & $\SI{-569}{}$  & $\SI{2410}{}$ & $\SI{-26300}{}$      \\ \hline
	$F_0$      & $\SI{1.23}{}$       & $\SI{1.23}{}$    & $\SI{1.28}{}$  & $\SI{1.06}{}$ & $\SI{6.81e-10}{}$      \\
	$g$       & $\SI{0.657}{}$       & $\SI{0.300}{}$    & $\SI{31.7}{}$  & $\SI{-28.8}{}$ & $\SI{-6480}{}$      \\
	\enddata
\end{deluxetable*}

\begin{deluxetable*}{cccccc} \label{tab:plasmaRateKIDAJ1}
	\tabletypesize{\small}
	\tablewidth{\textwidth}
	\tablecaption{Same as Table \ref{tab:plasmaRateKIDA} but for CH$^{+}$($ J=1 $). }
	\tablehead{
		\colhead{Parameter} & \multicolumn{5}{c}{Temperature range} \\  & \colhead{$ \SI[parse-numbers=false]{10-73.3}{K} $} & \colhead{$ \SI[parse-numbers=false]{73.3-681}{K} $} & \colhead{$ \SI[parse-numbers=false]{681-2680}{K} $} & \colhead{$ \SI[parse-numbers=false]{2680-18500}{K} $} & \colhead{$ \SI[parse-numbers=false]{18500-40000}{K} $}
	}
	\startdata
	$A$       & $\SI{1.84e-7}{}$       & $\SI{1.29e-7}{}$    & $\SI{2.36e-8}{}$  & $\SI{2.77e-7}{}$ & $\SI{4.51e-11}{}$      \\ 
	$\beta$       & $\SI{-0.846}{}$       & $\SI{-0.958}{}$    & $\SI{-2.69e-2}{}$  & $\SI{-0.632}{}$  &  $\SI{1.07}{}$   \\
	$\gamma$       & $\SI{2.18}{}$       & $\SI{-12.7}{}$    & $\SI{-646}{}$  & $\SI{2400}{}$ & $\SI{-29100}{}$      \\ \hline
	$F_0$      & $\SI{1.21}{}$       & $\SI{1.19}{}$    & $\SI{1.22}{}$  & $\SI{1.17}{}$ & $\SI{1.19e-11}{}$      \\
	$g$       & $\SI{0.866}{}$       & $\SI{2.35}{}$       & $\SI{15.2}{}$    & $\SI{-0.451}{}$  & $\SI{-7720}{}$       \\
	\enddata
\end{deluxetable*}

The second representation uses Arrhenius-Kooij functions, as are typically used in astrochemistry, combustion chemistry, and other chemical models and databases. However, with one such function, we cannot model all the experimentally observed features in the rate coefficient correctly. Instead, we provide a representation by $ m $ piecewise-joined Arrhenius-Kooij equations ($T_\mathrm{k}\in [T_{i},T_{i+1}] $, $ i\in \{0,...,m\} $) with
\begin{equation} \label{eq:plasmaRateFitKIDA}
	\alpha^{\mathrm{k}}_J(T_{\mathrm{k}})[\SI{}{cm^3\,s^{-1}}]=A_i\left(\frac{T_{\mathrm{k}}[\mathrm{K}]}{300}\right)^{\beta_i} e^{-\frac{\gamma_i}{T_{\mathrm{k}}[\mathrm{K}]}},
\end{equation}
where $ A_i $, $ \beta_i $, and $ \gamma_i $ are the fit parameters. Following KIDA conventions, the uncertainty is described by a log-normal factor $ F_J=\exp(\Delta \alpha^\mathrm{k}_{J}/\alpha^\mathrm{k}_{J}) $ for each interval with
\begin{equation} \label{eq:plasmaRateUncertaintyFitKIDA}
	F_J(T_{\mathrm{k}})=F_{0,i}\exp\left(g_i\left(\frac{1}{T_{\mathrm{k}}[\mathrm{K}]}-\frac{1}{300} \right)\right),
\end{equation}
where $ F_{0,i} $ and $ g_i $ are the fit parameters. We construct the data for $ F_J(T_{\mathrm{k}}) $ as average of $ F_{J,\mathrm{high}}(T_{\mathrm{k}})=\exp(\Delta_\mathrm{high} \alpha^\mathrm{k}_{J}/\alpha^\mathrm{k}_{J}) $ and $ F_{J,\mathrm{low}}(T_{\mathrm{k}})=\exp(\Delta_\mathrm{low} \alpha^\mathrm{k}_{J}/\alpha^\mathrm{k}_{J}) $, where $ \Delta_\mathrm{high} \alpha^\mathrm{k}_{J} $ and $ \Delta_\mathrm{low} \alpha^\mathrm{k}_{J} $ correspond to the upper and lower one-sigma errors, respectively (see Figure \ref{fig:plasmarate}(a)).

For astrochemical calculations, a continuous behavior of the functions at the interval borders is typically desired. However, this is not guaranteed in Equations (\ref{eq:plasmaRateFitKIDA}) and (\ref{eq:plasmaRateUncertaintyFitKIDA}). It can be shown that a continuous behavior of the fit functions can be achieved by introducing the following dependences of the $ A_i $ and $ F_{0,i} $ parameters 
\begin{eqnarray} 
	A_i&=&A_0\prod_{j=1}^{i}\left(\frac{T_j}{300}\right)^{\beta_{j-1}-\beta_j} e^{\frac{\gamma_j-\gamma_{j-1}}{T_j}},\label{eq:ParA}\\
	F_{0,i}&=&F_{0,0}\exp\left(\frac{g_i-g_0}{300}\right)\prod_{j=1}^{i}\exp\left(\frac{g_{j-1}-g_j}{T_j}\right).\qquad \label{eq:ParF}
\end{eqnarray}
Inserting Equations (\ref{eq:ParA}) and (\ref{eq:ParF}) into Equations (\ref{eq:plasmaRateFitKIDA}) and (\ref{eq:plasmaRateUncertaintyFitKIDA}), we obtain the continuous fit functions
\begin{eqnarray} \label{eq:plasmaRateFitKIDACont}
	\alpha^{\mathrm{k}}_J(T_{\mathrm{k}})[\SI{}{cm^3\,s^{-1}}]&=&A_0\left(\frac{T_{\mathrm{k}}[\mathrm{K}]}{300}\right)^{\beta_i}e^{-\frac{\gamma_i}{T_\mathrm{k}[\mathrm{K}]}}  \\ 
	&\times& \prod_{j=1}^{i}\left(\frac{T_j}{300}\right)^{\beta_{j-1}-\beta_j} e^{\frac{\gamma_j-\gamma_{j-1}}{T_j}},\nonumber
\end{eqnarray}
and
\begin{eqnarray} \label{eq:plasmaRateUncertaintyFitKIDACont}
	F_J(T_{\mathrm{k}})&=&F_{0,0}\exp\left(\frac{g_i-g_0}{300}\right)\exp\left(g_i\left(\frac{1}{T_{\mathrm{k}}[\mathrm{K}]}-\frac{1}{300} \right)\right)\nonumber\\
	&\times&\prod_{j=1}^{i}\exp\left(\frac{g_{j-1}-g_j}{T_j}\right),
\end{eqnarray}
for the rate coefficient and uncertainty fits, respectively. We note that for the first temperature interval ($i=0$) the product terms evaluate to unity and that the functions show the same dependence on $ T_{\mathrm{k}} $ as given by Equations (\ref{eq:plasmaRateFitKIDA}) and (\ref{eq:plasmaRateUncertaintyFitKIDA}), respectively.

For the fit with Equation (\ref{eq:plasmaRateFitKIDACont}), we let the interval borders $ T_i $ vary as free parameters and subsequently fixed them for the fit with Equation (\ref{eq:plasmaRateUncertaintyFitKIDACont}). The resulting fit parameters are used to calculate the  $ A_i $ and $ F_{0,i} $ parameters via Equations (\ref{eq:ParA}) and (\ref{eq:ParF}). All resulting parameters for Equations (\ref{eq:plasmaRateFitKIDA}) and (\ref{eq:plasmaRateUncertaintyFitKIDA}), using $ m=5 $ temperature intervals, are listed in Tables \ref{tab:plasmaRateKIDA} and \ref{tab:plasmaRateKIDAJ1} for the $J=0$ and $J=1$ levels, respectively. The relative deviations of the fit functions to the rate coefficients and error bounds are below $3.2\%$ at all temperatures.

The Arrhenius-Kooij representation of our rate coefficient in Equation (\ref{eq:plasmaRateFitKIDA}) suffers from two disadvantages. First, the piecewise function is not continuously differentiable. Second, for each feature in the rate coefficient, one interval has to be added to the function in order to model the data correctly. This introduces $ 3(N_r+1) $ fit parameters for $ N_r $ features. We recommend that chemical modelers use Equation (\ref{eq:plasmaRateFunction2}) to represent the kinetic temperature rate coefficient because it is continuously differentiable and requires fewer fit parameters.

\newpage

\bibliographystyle{aasjournal}
\bibliography{CH+Recombination.bib}




	
\end{document}